\documentclass[sigconf]{acmart}
\usepackage{amsmath}
\usepackage{multirow}
\usepackage{subfigure}

\usepackage{algorithm}
\usepackage{algorithmic}

\newcommand{\ci}{\perp\!\!\!\perp}

\newcommand{\algref}[1]{Algorithm~\ref{#1}}
\newcommand{\eqnref}[1]{Eq.~\eqref{#1}}
\newcommand{\figref}[1]{Figure~\ref{#1}}
\newcommand{\secref}[1]{Section~\ref{#1}}
\newcommand{\tableref}[1]{Table~\ref{#1}}
\newcommand{\thmref}[1]{Theorem~\ref{#1}}

\theoremstyle{remark}

\copyrightyear{2022}
\acmYear{2022}
\setcopyright{rightsretained}
\acmDOI{10.1145/3534678.3539041}

\acmConference[KDD '22] {Proceedings of the 28th ACM SIGKDD Conference on Knowledge Discovery and Data Mining}{August 14--18, 2022}{Washington, DC, USA}
\acmBooktitle{Proceedings of the 28th ACM SIGKDD Conference on Knowledge Discovery and Data Mining (KDD '22), August 14--18, 2022, Washington, DC, USA}
\acmPrice{}
\acmISBN{978-1-4503-9385-0/22/08}

\acmSubmissionID{apfp0357}
\begin{document}


% Keywords. The author(s) should pick words that accurately describe
% the work being presented. Separate the keywords with commas.

%%
%% The majority of ACM publications use numbered citations and
%% references.  The command \citestyle{authoryear} switches to the
%% If you are preparing content for an event
%% sponsored by ACM SIGGRAPH, you must use the "author year" style of
%% the next command will enable that style.

\title{Causal Inference-Based Root Cause Analysis for Online Service Systems with Intervention Recognition}

\author{Mingjie Li}
\orcid{0000-0002-4778-4098}
%\affiliation{
%	\institution{Tsinghua University}
%	\city{Beijing}
%	\country{China}
%}
%\email{lmj18@mails.tsinghua.edu.cn}

\author{Zeyan Li}
\orcid{0000-0002-3529-5879}
%\affiliation{
%	\institution{Tsinghua University}
%	\city{Beijing}
%	\country{China}
%}
%\email{zy-li18@mails.tsinghua.edu.cn}

\author{Kanglin Yin}
\orcid{0000-0002-9423-2085}
\affiliation{
	\institution{Tsinghua University}
	\city{Beijing}
	\country{China}
}
%\email{ykl_xsd@126.com}

\author{Xiaohui Nie}
\orcid{0000-0002-0371-854X}
%\affiliation{
%	\institution{BizSeer}
%	\city{Beijing}
%	\country{China}
%}
%\email{niexiaohui@bizseer.com}

\author{Wenchi Zhang}
\orcid{0000-0002-5599-030X}
%\affiliation{
%	\institution{BizSeer}
%	\city{Beijing}
%	\country{China}
%}
%\email{zhangwenchi@bizseer.com}

\author{Kaixin Sui}
\orcid{0000-0003-4545-7621}
\affiliation{
	\institution{BizSeer}
	\city{Beijing}
	\country{China}
}
%\email{suikaixin@bizseer.com}

\author{Dan Pei}
\authornote{
	Dan Pei is the corresponding author. Email: peidan@tsinghua.edu.cn
}
\orcid{0000-0002-5113-838X}
\affiliation{
	\institution{Tsinghua University}
	\city{Beijing}
	\country{China}
}
%\email{peidan@tsinghua.edu.cn}


\settopmatter{printacmref=true}
\fancyhead{}

% !TEX root=../main.tex
\begin{abstract}

Fault diagnosis is critical in many domains, as faults may lead to safety threats or economic losses.
In the field of online service systems, operators rely on enormous monitoring data to detect and mitigate failures.
Quickly recognizing a small set of root cause indicators for the underlying fault can save much time for failure mitigation.
In this paper, we formulate the root cause analysis problem as a new causal inference task named \textit{intervention recognition}.
We proposed a novel unsupervised causal inference-based method named \textit{Causal Inference-based Root Cause Analysis} (CIRCA).
The core idea is a sufficient condition for a monitoring variable to be a root cause indicator, \textit{i.e.}, the change of probability distribution conditioned on the parents in the Causal Bayesian Network (CBN).
Towards the application in online service systems, CIRCA constructs a graph among monitoring metrics based on the knowledge of system architecture and a set of causal assumptions.
% \wenchi{for root cause analysis problem}
The simulation study illustrates the theoretical reliability of CIRCA.
The performance on a real-world dataset further shows that CIRCA can improve the recall of the top-1 recommendation by 25\% over the best baseline method.

\end{abstract}

\maketitle

% !TEX root=../main.tex
\section{Introduction}\label{sec:introduction}

%Faults may bring economic or even health losses.
Fault diagnosis is critical in many domains, \textit{e.g.}, machinery maintenance~\cite{Yi:2021}, petroleum refining~\cite{Dhaou:2021}, and cloud system operations~\cite{Pool:2020,Zhang:2021},  which is an active research topic in the SIGKDD community.
In this work, we focus on root cause analysis (RCA) in online service systems (OSS), such as social networks, online shopping, search engine, \textit{etc}.
We adopt the terminology in \cite{Notaro:2021}, denoting a \textit{failure} as the undesired deviation in service delivery and a \textit{fault} as the cause of the failure.

With the expansion of system scale and the rise of microservice applications, OSS are more and more complex.
As a result, operators rely on monitoring data to understand what happens in the system~\cite{Beyer:2016}.
Common monitoring data include metrics, semi-structural logs, and invocation traces.
As the most widely available data, metrics are usually in the form of time series sampled at a constant frequency, \textit{e.g.}, once per minute.
Several metrics are the measures of the overall system health status, named the \textit{service level indicator}s (SLI), \textit{e.g.}, the average response time of an online service.
Once an SLI violates the pre-defined service level objective (\textit{i.e.}, a failure occurs), operators will mitigate the failure as soon as possible to prevent further damage.
As a single fault may propagate in the system~\cite{Gertler:1998} with multiple metrics being abnormal during a failure (named \textit{anomaly storm}~\cite{Zhao:2020}), RCA (recognizing a small set of \textit{root cause indicators}) of the underlying fault can save much time for failure mitigation.
% \wenchi{It's helpful to give examples of some common indicators, such as CPU utilization, service average response time and so on...}

With the rising emphasis on explainability in many domains, \textit{causal inference}~\cite{Yao:2021} has attracted much attention in the literature.
Though causal inference is promising, causal inference-based RCA is little studied, except Sage~\cite{Gan:2021} with counterfactual analysis.
In this paper, we novelly map a fault in OSS as an intervention~\cite{Pearl:2009} in causal inference.
From this point of view, we name a new causal inference task as \textit{intervention recognition} (IR), \textit{i.e.}, finding the underlying intervention based on the observations (Definition~\ref{def:intervention-recognition}).
Hence, we formulate RCA in OSS as an IR task.

The first challenge of the new IR task is the lack of a solution.
Though Sage~\cite{Gan:2021} conducts RCA via counterfactual analysis, the design of Sage implies an implicit assumption, \textit{i.e.}, there is no intervention to the system.
Hence, Sage is not a solution to the IR task.
Based on the definition of IR, we find that the probability distribution of an intervened variable changes conditioned on parents in the Causal Bayesian Network (CBN). This \nameref{thm:rc-criterion} points out an explainable way to conduct RCA.

The second challenge is to obtain the CBN for causal inference in OSS.
Many works have been done for \textit{causal discovery}~\cite{Guo:2020} from observational data.
MicroHECL~\cite{Liu:2021} and Sage~\cite{Gan:2021} utilize the call graph in OSS, which operators are familiar with.
However, these two works consider a few metrics, \textit{e.g.}, the latency between services.
We construct the CBN among metrics with the domain knowledge of system architecture, combined with a set of intuitive assumptions, handling more kinds of metrics than MicroHECL~\cite{Liu:2021} and Sage~\cite{Gan:2021}.

\begin{figure}[t]
	\includegraphics[width=0.85\columnwidth]{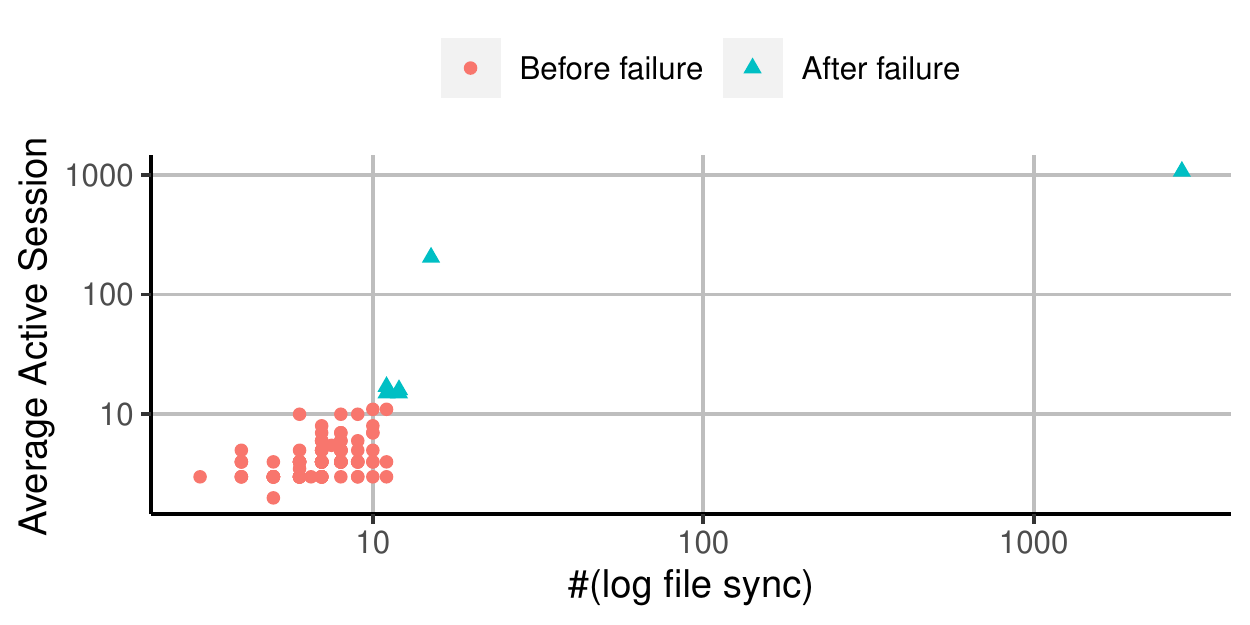}
	\caption{Joint distribution of the \textit{Average Active Session} (an SLI of the Oracle database) and the number of \textit{log file sync} waiting events within 2 hours. Each data point represents the two metrics' values at the same timestamp.}\label{fig:ood}
	\Description{
		This figure is a scatter plot between the number of \textit{log file sync} waiting events (x-Axis) and the \textit{Average Active Session} (AAS for y-Axis) around a high AAS failure.
		Observed data before the failure are in the bottom-left corner of the figure, while the data after the failure distribute towards the far top-right corner.
	}
\end{figure}

Thirdly, observational knowledge is incomplete, indicating the difficulty of reaching interventional knowledge even with a perfect CBN.
For example, \figref{fig:ood} shows the joint distribution of the \textit{Average Active Session} (AAS) and the number of \textit{log file sync} waiting events around a high AAS failure of an Oracle database instance.
% Observed data before the failure are in the bottom-left corner of the figure, while the data after the failure distribute towards the far top-right corner.
% The lack of overlap blocks the comparison between these two distributions for intervention recognition.
Observed data before the failure are in the bottom-left corner of the figure.
Hence, how AAS normally distributes is missing when ``\#(log file sync)'' is larger than 1,000, where the data after the failure distribute.
The lack of overlap between the two distributions around the failure blocks recognizing intervention, if any, in AAS.
To address this challenge, we transform distribution comparison as point-wise hypothesis testing via the regression technique.
Moreover, a descendant adjustment technique is proposed to alleviate the bias introduced by a poor understanding of the system's normal status in the hypothesis testing.

We implement the proposed Causal Inference-based Root Cause Analysis (CIRCA).
CIRCA outperforms baseline methods in our simulation study, illustrating its theoretical reliability.
We further evaluate CIRCA with a real-world dataset.
CIRCA improves the recall of the top-1 recommendation by 25\% over the best baseline method, which shows the practical potential of our approach.
The contributions of this work are summarized as follows.

\begin{itemize}
	\item For the first time in the literature, we formulate the RCA problem in OSS as a new causal inference task named \textit{intervention recognition} (Definition~\ref{def:intervention-recognition}).
	Utilizing the advance of causal inference, we find a practical criterion to locate the root cause (\thmref{thm:rc-criterion}).
% 	Utilizing the advance of causal inference, we find a criterion to locate the root cause (\thmref{thm:rc-criterion}), which is proven to be practical.
	\item We propose Causal Inference-based Root Cause Analysis (CIRCA) for OSS.
	We propose a practical guideline to construct the CBN with the knowledge of system architecture.
	Two more techniques, namely regression-based hypothesis testing and descendant adjustment, are proposed to infer root cause metrics in the graph.
	\item CIRCA is evaluated with both simulation and real-world datasets.
	The simulation study illustrates CIRCA's theoretical reliability, while the real-world dataset shows CIRCA's practical value over baseline methods.
\end{itemize}

% The rest of this paper is organized as follows.
% In the next section, we first provide preliminary material, followed by the formal definitions of IR and RCA discussed in this work.
% \secref{sec:rc-criterion} introduces our analysis framework, while \secref{sec:methodology} describes the design of CIRCA.
% In \secref{sec:experiments}, We evaluate different methods on both simulation and real-world datasets.
% Finally, we introduce related works in \secref{sec:related-work} and conclude in \secref{sec:conclusion}.

% !TEX root=../main.tex
\section{Problem Formulation}

\subsection{Preliminary}

\noindent \textbf{Notation.}
% Throughout this paper, we use the following conventions.
An upper case letter (\textit{e.g.}, $X$) refers to a variable (metric), while a lower case letter (\textit{e.g.}, $x$) represents an \textit{assignment} of the corresponding variable.
By \textit{assignment}, we mean one of the possible values. 
To distinguish variables (values) at different times in a time series, the timestamp will be put on the letter as a superscript.
For example, denote AAS as an upper case letter $Y$, and $y^{(t)}$ refers to the value of AAS at time $t$.
Denote the value range of $Y$ as $Val(Y)$, then we have $y^{(t)} \in Val(Y) = \{0\} \cup \mathbb{R}^{+}$ for the non-negative numeric AAS.
A boldfaced letter means a set of elements (variables or values), \textit{e.g.}, we denote all the metrics as $\mathbf{V}$ while $\mathbf{v}$ is an assignment of $\mathbf{V}$.

% Throughout this work, we use the following conventions.
% An upper case letter (\textit{e.g.}, $X$) refers to a variable (a metric), while a lower case letter (\textit{e.g.}, $x$) represents an assignment of the corresponding variable.
% A bold-face letter means a set of elements, \textit{e.g.}, we denote all the metrics as $\mathbf{V}$ while $\mathbf{v}$ is an assignment of $\mathbf{V}$.
% Possible values of $\mathbf{X}$ are written as $Val(\mathbf{X})$, \textit{i.e.}, $\mathbf{x} \in Val(\mathbf{X})$.
% To distinguish variables (values) in a time-series, the timestamp will be put as a superscript, \textit{e.g.}, $x^{(t)} \in Val^{(t)}(X)$.

\noindent \textbf{The Ladder of Causation.}
% Judea Pearl discovers and studies the ``Ladder of Causation''~\cite{Pearl:2018}.
We formulate the problem with Judea Pearl's ``Ladder of Causation''~\cite{Bareinboim:2022}.
The first layer of the causal ladder encodes the observational knowledge $\mathcal{L}_{1}(\mathbf{V}) = P(\mathbf{V})$, where $P(\mathbf{V})$ is a joint probability distribution.
% \ykl{, where P(V) means ....}
% Meanwhile, the second layer encodes the interventional knowledge $\mathcal{L}_{2}(\mathbf{V}, \mathbf{m}) = P_{\mathbf{m}}(\mathbf{V})$, where $P_{\mathbf{m}}(\mathbf{V}) = P(\mathbf{V} \mid do(\mathbf{m}))$ and $\mathbf{M} \subseteq\mathbf{V}$.
Meanwhile, the second layer encodes the interventional knowledge $\mathcal{L}_{2}(\mathbf{m}) = P_{\mathbf{m}}$, where $P_{\mathbf{m}}(\mathbf{V}) = P(\mathbf{V} \mid do(\mathbf{m}))$ and $\mathbf{M} \subseteq\mathbf{V}$.
% $P(\mathbf{V} \mid do(\mathbf{m}))$ describes the probability distribution of $\mathbf{V}$ when $\mathbf{M}$ is fixed to the given value $\mathbf{m}$, where $do$ is named \textit{do-operator}~\cite{Pearl:2009}, representing an intervention.
The \textit{do-operator} $do(\mathbf{m})$ means fixing variables $\mathbf{M}$ to the given values $\mathbf{m}$, also called an \textit{intervention}~\cite{Pearl:2009}.
So that $P(\mathbf{V} \mid do(\mathbf{m}))$ indicates the probability distribution over $\mathbf{V}$ under the intervention to $\mathbf{M}$.
Finally, the third layer encodes the counterfactual knowledge, reasoning about what if another situation happened in the past.
For example, it requires the counterfactual knowledge to predict the latency with sufficient computing resources when high latency and full CPU usage are observed.
The hierarchy of the causal ladder \textit{almost never} collapses (named CHT, Causal Hierarchy Theorem~\cite{Bareinboim:2022}).
If we want to answer the question at Layer i, we need knowledge at Layer i or higher~\cite{Bareinboim:2022}.
% The ladder of causation defines a hierarchy that \textit{almost never} collapses, named Causal Hierarchy Theorem (CHT)~\cite{Bareinboim:2022}.
% In other words, ``to answer questions at Layer i, one needs knowledge at Layer i or higher.''~\cite{Bareinboim:2022}
% \ykl{The hierarchy of the Ladder of Causation(named CHT, Causal Hierarchy Theorem ~\cite{Bareinboim:2022}) \textit{almost never} collapses. When we want to answer the question at Layer i, we have to needs knowledge at Layer i or higher.~\cite{Bareinboim:2022}}

\noindent \textbf{Structural Causal Model (SCM).}
We model the relations among metrics via the structural causal model~\cite{Pearl:2009}.
An SCM contains a set of structural equations shown in \eqnref{eqn:scm}, where $V_{i} \in \mathbf{V}$ and $\mathbf{Pa}(V_{i}) \subseteq \mathbf{V}$.
% The parameters in \eqnref{eqn:scm} can be divided into two parts: 1) observed variables $\mathbf{Pa}(V_{i})$, named parents (direct causes) of $V_{i}$, and 2) unobserved ones $\mathbf{u}_i$, \textit{e.g.}, noises that we are not interested in.
\eqnref{eqn:scm} contains two kinds of parameters:
1) assignments of observed variables $\mathbf{Pa}(V_{i})$, named parents (direct causes) of $V_{i}$, and
2) assignments of unobserved variables $\mathbf{U}_i$, where $\mathbf{U}_{i} \cap \mathbf{V} = \emptyset$.

\begin{equation}\label{eqn:scm}
	v_{i} = f_{i}(\mathbf{pa}(V_{i}), \mathbf{u}_{i})
\end{equation}

Denote the graph encoded by the SCM as $\mathcal{G} = (\mathbf{V}, \mathbf{E})$, where $\mathbf{E} = \{V_{j} \to V_{i} \mid V_{j} \in \mathbf{Pa}(V_{i})\}$ is the set of directed edges.
In contrast to $\mathbf{Pa}$, $\mathbf{Ch}(V_{i}) = \{V_{j} \mid V_{i} \in \mathbf{Pa}(V_{j})\}$ represents the children of $V_{i}$.
This work rests on the following assumptions.
\begin{description}
    \item[DAG] $\mathcal{G}$ is a directed acyclic graph (DAG)~\cite{Pearl:2009}, following related works in OSS~\cite{Chen:2014,Wang:2018,Gan:2021}.
    \item[Markovian] ``The exogenous parent sets $\mathbf{U}_{i}, \mathbf{U}_{j}$ are independent whenever $i \neq j$''~\cite{Bareinboim:2022}, \textit{i.e.}, $(\forall i \neq j) \mathbf{U}_{i} \ci \mathbf{U}_{j}$, where $\ci$ means independent.
    % \item[Causal Sufficiency] ``All common drivers are among the observed variables''~\cite{Runge:2019}, \textit{i.e.}, $(\forall i \neq j) \mathbf{U}_{i} \ci \mathbf{U}_{j}$, where $\ci$ means independent.
    % \item[Causal Sufficiency] The SCM under discussion is \textit{Markovian}~\cite{Bareinboim:2022} (also known as the \textit{Causal Sufficiency} assumption~\cite{Runge:2019}), \textit{i.e.}, $(\forall i \neq j) \mathbf{U}_{i} \ci \mathbf{U}_{j}$, where $\ci$ means independent;
    \item[Faithfulness~\cite{Pearl:2009}] Any intervention makes an observable change, \textit{i.e.}, $P(V_{i} \mid \mathbf{pa}(V_{i}), do(v_{i})) \neq P(V_{i} \mid \mathbf{pa}(V_{i}))$.
\end{description}
Under the DAG assumption and the Markovian assumption, $\mathcal{G}$ can be taken as a CBN~\cite{Bareinboim:2022}.

\subsection{Root Cause Analysis and Causal Inference}

We set up a concept mapping between the RCA problem and causal inference.
\begin{itemize}
    \item A fault in OSS is mapped to an unexpected intervention;
    \item Fault-free data come from the observational distribution;
    \item Faulty data come from an interventional distribution.
\end{itemize}
Based on the mapping above, we define a new causal inference task as \textit{intervention recognition} (Definition~\ref{def:intervention-recognition}).
We formulate RCA discussed in this work as an intervention recognition task in OSS.

\begin{definition}[Intervention Recognition, IR]\label{def:intervention-recognition}
    For a given SCM $\mathcal{M}$, let $\mathcal{L}_{1}$ be the observational distribution of $\mathcal{M}$ and $P_{m} = P(\mathbf{V} \mid do(\mathbf{m}))$ be the interventional distribution of a certain intervention $do(\mathbf{m})$.
    \textit{Intervention recognition} is to find $\mathbf{m}$ based on $\mathcal{L}_{1}$ and $P_{m}$.
\end{definition}

\begin{definition}[Root Cause]\label{def:root-cause}
    The \textit{root cause} is the intervened variables ($\mathbf{M}$).
    Each element of $\mathbf{M}$ is named a \textit{root cause variable}.\footnote{We also use \textit{root cause indicator} and \textit{root cause metric} according to the context.}
\end{definition}

\section{Intervention Recognition Criterion}\label{sec:rc-criterion}

% 明杰：
%   我觉得“argue”比“prove”更合适
%   Theorem 3.1的成立是有条件的，正文中的这句话没有交代条件
%   把条件写在正文里会更难懂
We argue that IR shall be positioned at the second layer in the ladder of causation, as shown in \thmref{thm:rca-l2}.
The proof of \thmref{thm:rca-l2} is provided in Appendix~\ref{sec:proof-rca-l2}.
The key to the proof is that IR is the inverse mapping of $\mathcal{L}_{2}$ under the adopted assumptions.
Combining \thmref{thm:rca-l2} with CHT~\cite{Bareinboim:2022}, we further obtain Corollary~\ref{thm:l2-necessary} and \ref{thm:l3-unnecessary}.

\begin{theorem}\label{thm:rca-l2}
	For a given SCM $\mathcal{M}$ with a CBN $\mathcal{G}$, the knowledge of IR for $\mathcal{M}$ is equivalent to $\mathcal{L}_{2}$ under the Faithfulness assumption.
\end{theorem}

\begin{corollary}\label{thm:l2-necessary}
	We need the knowledge at Layer 2 (interventional) to conduct IR.
\end{corollary}

\begin{corollary}\label{thm:l3-unnecessary}
	The knowledge at Layer 3 (counterfactual) is not necessary to conduct IR.
\end{corollary}

Hence, we propose to take full advantage of the CBN, as the CBN is a known bridge between observational data and interventional knowledge~\cite{Bareinboim:2022}.
We argue that \thmref{thm:rc-criterion} is a necessary and sufficient condition for a variable to be intervened.
The proof of \thmref{thm:rc-criterion} is provided in Appendix~\ref{sec:proof-rc-criterion}.
Based on our concept mapping between RCA and causal inference, the \nameref{thm:rc-criterion} is also a criterion to find root cause indicators.

\begin{theorem}[Intervention Recognition Criterion]\label{thm:rc-criterion}
	Let $\mathcal{G}$ be a CBN and $\mathbf{Pa}(V_{i})$ be the parents of $V_{i}$ in $\mathcal{G}$.
	Under the Faithfulness assumption, $V_{i}$ is intervened \textit{iff} $V_{i}$ no longer follows the distribution defined by $\mathbf{pa}(V_{i})$, \textit{i.e.},
	\begin{equation*}
		V_{i} \in \mathbf{M} \Leftrightarrow P_{\mathbf{m}}(V_{i} \mid \mathbf{pa}(V_{i})) \neq \mathcal{L}_{1}(V_{i} \mid \mathbf{pa}(V_{i}))
	\end{equation*}
\end{theorem}

% When a fault occurs and affects $V_{i}$ directly (\textit{i.e.}, $V_{i} \in \mathbf{M}$), $v_{i}$ is no longer defined by $f_{i}$ in \eqnref{eqn:scm}\footnote{We refer readers to the Manipulation Theorem~\cite{Spirtes:1993} for more discussion.}.
% There are only two basic situations:
% 1) $f_{i}$ is replaced by $f_{i}^{\prime}$, or
% 2) the distribution of unobserved variables changed from $P(\mathbf{U}_{i})$ to $P^{\prime}(\mathbf{U}_{i})$.
% It is a special case of the first situation that $\mathbf{Pa}(V_{i})$ (or $\mathbf{U}_{i}$) is changed to $\mathbf{Pa}^{\prime}(V_{i})$ (or $\mathbf{U}_{i}^{\prime}$).
% If only $\mathbf{pa}(V_{i})$ (the value of $\mathbf{Pa}(V_{i})$ ) changes, $V_{i}$ should not be considered for the root cause, as the failure just propagates through $V_{i}$.
% In contrast, $V_{i}$ is the only visible indicator of $\mathbf{U}_{i}$ under the assumption of Causal Sufficiency.
% Hence, $V_{i}$ is the desired result if $\mathbf{u}_{i}$ changes.
% In summary, we argue that the following criterion (\thmref{thm:rc-criterion}) is a necessary and sufficient condition for a metric to be an indicator of the root cause.
% The proof of \thmref{thm:rc-criterion} is provided in Appendix~\ref{sec:proof-rc-criterion}.

% !TEX root=../main.tex
\section{Causal Inference-Based Root Cause Analysis}\label{sec:methodology}

In this section, we propose a novel method named \textit{CIRCA}.
We first present a structural way to determine the parents $\mathbf{Pa}(V_{i})$ for each metric $V_{i}$ based on system architecture.
CIRCA adopts \textit{regression-based hypothesis testing} (RHT) to deal with the incomplete distribution of faulty data.
To address the challenge of the incomplete distribution of fault-free data, CIRCA adjusts the anomaly score for suspicious metrics based on their descendants in the CBN.

\subsection{Structural Graph Construction}

We propose the \textit{structural graph} (SG) as the CBN for OSS.
SG combines the system architecture knowledge with a set of assumptions, which may not suit domains other than OSS.
We first classify monitoring metrics into four dimensions, named \textit{meta metric}s.
Several causal assumptions among those four kinds of meta metrics provide the building blocks of an SG.
% We first recognize four measurement dimensions (\textit{meta metric}s) of a computing process.
% The assumptions among those four kinds of meta metrics provide the building blocks of an SG.
We further extend the system with the architecture of components to construct a graph at the meta metric level, named a \textit{skeleton}.
Finally, we plug monitoring metrics into the corresponding meta metric to obtain the SG.
% Finally, we fulfill the skeleton with monitoring metrics to obtain the SG, replacing the corresponding meta metric.
\algref{alg:structural-graph-construction} summarizes the overall procedure.

\subsubsection{Meta Metrics}

In general, a service takes \textit{input} and produces \textit{output}.
Each request lasts for some \textit{time} and consumes some \textit{resources}.
We take those dimensions as four \textit{meta metrics} of a service, named after the four golden signals in site reliability engineering~\cite{Beyer:2016}.
\textit{Traffic}, \textit{Errors}, and \textit{Latency} measure the distribution of input, output, and processing time, respectively.
We classify other monitoring metrics as resource consumption, denoted as \textit{Saturation}.
% For example, the CPU usage will be classified as Saturation, describing a resource (the CPU device).

\begin{figure}[htb]
	\subfigure[Causal assumptions within a service]{\label{fig:tsel}
		\centering
		\includegraphics{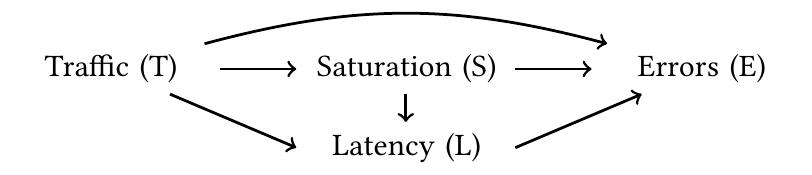}
	} \\
	\subfigure[
	    The skeleton of one web service (WEB) with its dependent database (DB).
	    We plug DB's meta metrics into the \textit{Saturation} of WEB.
	]{\label{fig:tsel-extend}
		\centering
		\includegraphics{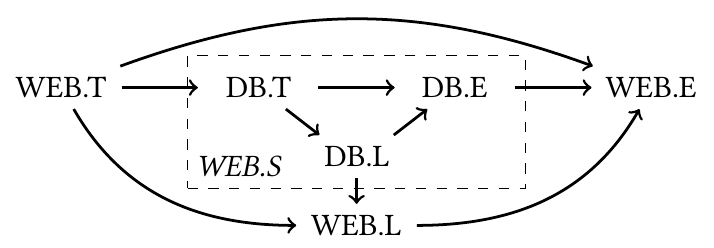}
	}
% Following is compatible to the package subcaption
% 	\begin{subfigure}{\columnwidth}
% 		\centering
% 		\includegraphics{img/tsel}
% 		\caption{Causal assumptions within a service}\label{fig:tsel}
% % 		\caption{within a web service}\label{fig:tsel}
% 	\end{subfigure}

% 	\begin{subfigure}{\columnwidth}
% 		\centering
% 		\includegraphics{img/tsel-extend}
% 		\caption{
% 		    The skeleton of one web service (WEB) with its dependent database (DB).
% 		    We plug DB's meta metrics into the \textit{Saturation} of WEB.
% 		}\label{fig:tsel-extend}
% % 		\caption{between one web service (WEB) and its dependency database(DB), fulfilling a resource (DB) with its meta metrics}\label{fig:tsel-extend}
% 	\end{subfigure}
	\caption{Causal assumptions among meta metrics}
	\Description{
		\textit{Traffic} is assumed as the cause of \textit{Saturation}, \textit{Latency}, and \textit{Errors}.
		\textit{Saturation} is assumed as the cause of \textit{Latency} and \textit{Errors}.
		\textit{Latency} is assumed as the cause of \textit{Errors}.
	}
\end{figure}

We assign directions for the relations among these four meta metrics in \figref{fig:tsel}.
As the start of a request, \textit{Traffic} is assumed to be the cause of all other three meta metrics, while \textit{Errors} (the end of a request) are taken as the effect of others.
% The edge from \textit{Saturation} to \textit{Latency} encodes the queuing phenomenon, \textit{i.e.}, resource consumption will not postpone a request until some kinds of resources become the bottleneck.
The edge from \textit{Saturation} to \textit{Latency} encodes our preference on the former, as resource consumption is one of the common considerations for large latency in OSS~\cite{Gan:2021}.

\subsubsection{Skeleton with Architecture Extension}

A complex OSS system will invoke multiple services to process one single request.
Meanwhile, there will be multiple components for monolithic OSS.
% or instructions for a function
Based on the architecture knowledge encoded in the call graph, we construct the \textit{skeleton} among meta metrics of the system and all its dependent services.
For a web service (WEB) and its database (DB) shown in \figref{fig:tsel-extend}, we take DB as a resource of WEB.
The part of WEB's Saturation that measures DB will be extended into DB's meta metrics, which inherit the relations between WEB's Saturation and other meta metrics of WEB.
The extension will be applied to each service in the call graph.
In summary, we introduce three more causal assumptions between a service and its dependent ones.

\begin{itemize}
	\item The caller's \textit{Traffic} influences the callees' \textit{Traffic};
	\item The callees' \textit{Latency} contributes to the caller's \textit{Latency};
	\item The caller's output is calculated based on the callees' output.
\end{itemize}

\subsubsection{Monitoring Metric Plugging-in}

Finally, we plug monitoring metrics in meta metrics to obtain the SG.
A mapping is required to describe which dimension of which service each monitoring metric measures.
There can be some meta metrics that do not have any monitoring metrics.
For example, the common measurement for memory is just usage (\textit{Traffic}), while the speed (\textit{Latency}) is unavailable.
Moreover, one monitoring metric can be derived from multiple meta metrics.
For example, \textit{DB access per request} is calculated by the Traffic of both a web service and a database.

\algref{alg:structural-graph-construction} describes the plugging-in process after skeleton construction.
SG links monitoring metrics from one meta metric to its children (Line \ref{line:add-edges-parent2child}).
Monitoring metrics that are derived from multiple meta metrics may introduce self-loop.
To avoid such cycles, the monitoring metric for the last meta metric in topological order will be taken as the common effect of other meta metrics (from Line \ref{line:deal-with-multi-corresponding:start} to Line \ref{line:deal-with-multi-corresponding:end}).
Moreover, meta metrics measuring the dimension of \textit{Errors} will be accumulated for descendants (Line \ref{line:accumulate-error}), as broken data may not be validated in time.

During the process, an empty meta metric will gather the monitoring metrics of its parents for its children (Line \ref{line:gather-metrics-for-children}).
Consider a meta metric ($V_{i}^{m}$), one of its parents without monitoring ($V_{j}^{m}$), and their structural equations ($f_{i}^{m}$ and $f_{j}^{m}$).
We can substitute $f_{j}^{m}$ for unobserved $V_{j}^{m}$ in $f_{i}^{m}$, as shown in \eqnref{eqn:scm-substitute-variable}.
Both the parents of $V_{j}^{m}$ and those of $V_{i}^{m}$ (except $V_{j}^{m}$) show as the parameters of $f_{i}^{m \prime}$, which is the reason behind Line \ref{line:gather-metrics-for-children}.

\begin{equation}\label{eqn:scm-substitute-variable}
	\begin{aligned}
		v_{i}^{m} & = f_{i}^{m}\left(
		f_{j}^{m}\left(\mathbf{pa}_{\mathcal{G}_{skel}}(V_{j}^{m}), \mathbf{u}_{j}^{m}\right),
		\mathbf{pa}_{\mathcal{G}_{skel}}(V_{i}^{m}) \setminus \{v_{j}^{m}\}, \mathbf{u}_{i}^{m}
		\right) \\
		& = f_{i}^{m \prime}\left(
		\mathbf{pa}_{\mathcal{G}_{skel}}(V_{j}^{m}),
		\mathbf{pa}_{\mathcal{G}_{skel}}(V_{i}^{m}) \setminus \{v_{j}^{m}\},
		\mathbf{u}_{i}^{m}, \mathbf{u}_{j}^{m}
		\right)
	\end{aligned}
\end{equation}

\begin{algorithm}[tb]
	\caption{Structural Graph Construction}\label{alg:structural-graph-construction}
	\begin{algorithmic}[1]
		\REQUIRE $\mathcal{G}_{c}$, the call graph; $\mathbf{h} : \mathbf{V}^{m} \to 2^{\mathbf{V}}$, the mapping from meta metrics $\mathbf{V}^{m}$ to monitoring metrics
		\STATE $\mathcal{G}_{s} \gets$ initial the structure graph
%		\STATE $\mathbf{h}^{\prime}: \mathbf{V}^{m} \to 2^{\mathbf{V}}$ will be an updated mapping
%		\STATE $\mathbf{r}: \mathbf{V} \to 2^{\mathbf{V}^{m}}$ records the references $\{V_{j}^{m} \mid V_{i} \in \mathbf{h}(V_{j}^{m})\}$ for each monitoring metric $V_{i}$
		\STATE $\mathcal{G}_{skel} \gets$ construct the skeleton based on $\mathcal{G}_{c}$
		\FORALL{$V_{i}^{m} \in \mathbf{V}^{m}$ in a topological order from $\{V_{j}^{m} \mid \lvert \mathbf{Pa}_{\mathcal{G}_{skel} }(V_{j}^{m}) \rvert = 0 \}$}
%			\STATE{$\mathbf{Q}_{i} \gets \bigcup_{V_{j}^{m} \in \mathbf{Pa}_{\mathcal{G}_{skel} }(V_{i}^{m})}\mathbf{h}^{\prime}(V_{j}^{m})$}
			\STATE{$\mathbf{Q}_{i} \gets$ Collect monitoring metrics of $\mathbf{Pa}_{\mathcal{G}_{skel} }(V_{j}^{m})$}
%			\STATE{$\mathbf{C}_{i} \gets \mathbf{h}(V_{i}^{m})$}
			\STATE{$\mathbf{C}_{i} \gets$ Collect monitoring metrics of $V_{i}^{m}$}
			\FOR{$V_{j} \in \mathbf{h}(V_{i}^{m})$}\label{line:deal-with-multi-corresponding:start}
%				\IF{$\lvert \mathbf{r}(V_{j}) \rvert > 1$}
				\IF{$V_{j}$ is mapped to multiple meta metrics}
					\STATE{$\mathbf{C}_{i} \gets \mathbf{C}_{i} \setminus \{V_{j}\}$}
					\COMMENT{Prevent self loop of $V_{j}$}
%					\IF{$V_{i}^{m}$ is the last meta metric in $\mathbf{r}(V_{j})$}
					\IF{$V_{j}$ is visited for the last time}
%						\STATE{Add edges from $\bigcup_{V_{k}^{m} \in r(V_{j}), i \neq k}\mathbf{h}^{\prime}(V_{k}^{m})$ to $V_{j}$ in $\mathcal{G}_{s}$}
						\STATE{Add edges from the corresponding meta metrics of $V_{j}$ other than $V_{i}^{m}$ to $V_{j}$ in $\mathcal{G}_{s}$}
						\STATE{$\mathbf{Q}_{i} \gets \mathbf{Q}_{i} \cup \{V_{j}\}$}
						\COMMENT{Take it as the proxy of others}
					\ENDIF
				\ENDIF
			\ENDFOR\label{line:deal-with-multi-corresponding:end}
			\COMMENT{Deal with monitoring metrics that are derived from multiple meta metrics}
			\STATE{Add edges from $\mathbf{Q}_{i}$ to $\mathbf{C}_{i}$ in $\mathcal{G}_{s}$}\label{line:add-edges-parent2child}
			\IF{$V_{i}^{m}$ represents Errors}
%				\STATE{$\mathbf{C}_{i} \gets \mathbf{C}_{i} \cup \bigcup_{V_{j}^{m}}\mathbf{h}^{\prime}(V_{j}^{m})$, where $V_{j}^{m} \in \mathbf{Pa}_{\mathcal{G}_{skel} }(V_{i}^{m})$ and $V_{j}^{m}$ also represents Errors}\label{line:accumulate-error}
				\STATE{Update $\mathbf{C}_{i}$ with monitoring metrics from the Errors-representing meta metrics in $\mathbf{Pa}_{\mathcal{G}_{skel} }(V_{j}^{m})$}\label{line:accumulate-error}
			\ENDIF
			\COMMENT{Transfer Errors}
			\IF{$\mathbf{C}_{i}  = \emptyset$}
%				\STATE{$\mathbf{h}^{\prime}(V_{i}^{m}) \gets \mathbf{Q}_{i}$}\label{line:gather-metrics-for-children}
				\STATE{$\mathbf{h}(V_{i}^{m}) \gets \mathbf{Q}_{i}$}\label{line:gather-metrics-for-children}
				\COMMENT{Gather monitoring metrics for children}
			\ELSE
%				\STATE{$\mathbf{h}^{\prime}(V_{i}^{m}) \gets \mathbf{C}_{i}$}
				\STATE{$\mathbf{h}(V_{i}^{m}) \gets \mathbf{C}_{i}$}
			\ENDIF
		\ENDFOR
		\RETURN $\mathcal{G}_{s}$
	\end{algorithmic}
\end{algorithm}

\subsection{Regression-based Hypothesis Testing}

% The understanding of $P_{\mathbf{m}}$ is limited due to the requirement to mitigate the failure as soon as possible.
The understanding of $P_{\mathbf{m}}$ is restricted by mitigating the failure as soon as possible.
Instead of comparing two distributions directly, we reformulate the \nameref{thm:rc-criterion} as hypothesis testing with the following null hypothesis ($\mathbf{H}_{0}$) for each metric $V_{i}$.
% Whether the value comes from an expected distribution can be measured by probability density (probability) for a continuous (discrete) metric.
% We should reject $\mathbf{H}_{0}$ if the probability density (probability) of the observation is zero.

\begin{description}
	\item[$\mathbf{H}_{0}$]  $V_{i}$ is not an indicator of the root cause, \textit{i.e.}, $$V_{i}^{(t)} \sim \mathcal{L}_{1}\left(V_{i}^{(t)} \mid \mathbf{pa}^{(t)}(V_{i})\right)$$
\end{description}

We utilize the regression technique to calculate the expected distribution $\mathcal{L}_{1}\left(V_{i}^{(t)} \mid \mathbf{pa}^{(t)}(V_{i})\right)$.
A regression model is trained for each variable with data before the fault is detected, performing as a proxy of the corresponding structural equation.
Let $\bar{v}_{i}^{(t)}$ be the regression value for $v_{i}^{(t)}$.
Assuming that the residuals follow an \textit{i.i.d.} normal distribution $N(\mu_{\epsilon,i}, \sigma_{\epsilon,i})$, \eqnref{eqn:anomaly-score:point} measures to what extent a new datum $v_{i}^{(t)}$ deviates from the expected distribution, denoted as $a_{V_{i}}^{(t)}$.
\eqnref{eqn:anomaly-score} further aggregates $a_{V_{i}}^{(t)}$ for all the available data during the abnormal period as the anomaly score of $V_{i}$.

\begin{equation}\label{eqn:anomaly-score:point}
	a_{V_{i}}^{(t)} = \left|\frac{\left(v_{i}^{(t)} - \bar{v}_{i}^{(t)}\right) - \mu_{\epsilon,i}}{\sigma_{\epsilon,i}}\right|
\end{equation}

\begin{equation}\label{eqn:anomaly-score}
	s_{V_{i}} = \max_{t} a_{V_{i}}^{(t)}
\end{equation}

\subsection{Descendant Adjustment}
% \nxh{better clarify the logic of this paragraph.}
There will be bias in the regression results due to a poor understanding of $\mathcal{L}_{1}$. 
% Inspired by the common queuing phenomenon in an online service system, we adjust the anomaly score of one metric with those of its descendants.
% For example, reducing traffic or supplementing resources is actionable to restore the low latency.
% Hence, we prefer to assign higher scores for traffic and resource utilization (the parents of latency in the CBN) than latency's score.
We adjust the anomaly score of one metric with those of its descendants.
Our intuition is that when both a metric and one of its parents in the CBN is abnormal, we prefer the latter.
For example, supplementing extra resources is an actionable mitigation method to restore the low latency.
Hence, we assign a higher score for resource utilization (the parents of latency in the CBN) than latency's score.

We summarize the adjustment in \algref{alg:descendant-adjaustment}.
The children of a metric $V_{i}$ are first considered (Line \ref{line:children:adjustment}).
We exclude some metrics ($\{V_{i} \mid s_{V_{i}} < 3\}$) from the root cause indicators, so called the \textit{three-sigma rule of thumb}.
As the failure propagates through them, those metrics will gather anomaly scores from children for the candidate root cause in their ancestors (Line \ref{line:children:propagate}).
Finally, the anomaly score of $V_{i}$ ($s_{V_{i}}$) will increase by the maximum of descendants' scores just mentioned (Line \ref{line:children:aggregate}).

\begin{algorithm}[tb]
	\caption{Descendant Adjustment}\label{alg:descendant-adjaustment}
	\begin{algorithmic}[1]
		\REQUIRE $\mathbf{s}$, anomaly scores by \eqnref{eqn:anomaly-score}
% 			$\tau \gets 3$, the threshold to estimate whether the failure just propagates through a metric
		\STATE $S \gets$ a mapping from $V_{i}$ to the anomaly scores $S(V_{i})$ that may be the direct effect of $V_{i}$
		\FOR{$V_{i} \in \mathbf{V}$ in a topological order from $\{V_{j} \mid \lvert \mathbf{Ch}(V_{j}) \rvert = 0$\}}
			\STATE $S(V_{i}) \gets \{s_{V_{j}} \mid V_{j} \in \mathbf{Ch}(V_{i})\}$\label{line:children:adjustment}
			\FOR{$V_{j} \in \mathbf{Ch}(V_{i})$}
				\IF{$s_{V_{j}} < 3$}
					\STATE{$S(V_{i}) \gets S(V_{i}) \cup S(V_{j})$}\label{line:children:propagate}
				\ENDIF
			\ENDFOR
		\ENDFOR
		\COMMENT{Collect direct effects}
		\FOR{$V_{i} \in \mathbf{V}$}
			\IF{$s_{V_{j}} \ge 3$}
				\STATE{$s_{V_{i}}^{\prime} \gets s_{V_{i}} + \max( S(V_{i}) )$}\label{line:children:aggregate}
				\COMMENT{Adjust based on descendants}
			\ENDIF
		\ENDFOR
		\RETURN $\mathbf{s}^{\prime}$, the adjusted anomaly scores
	\end{algorithmic}
\end{algorithm}

% !TEX root=../main.tex
\section{Experiments}\label{sec:experiments}

In this section, we compare the performance of different methods.
We first conduct a simulation study to verify their theoretical reliability.
The effectiveness is further evaluated on a real-world dataset.
All the execution duration is measured on a server with an Intel Xeon E5-2620 CPU @ 2.40GHz (22 cores) and 57GB RAM.
We release our code at \url{https://github.com/NetManAIOps/CIRCA}.
More experiment details are in Appendix~\ref{sec:implementation-details}. 

\subsection{Experimental Setup}

\subsubsection{Hyperparameters}

The shortest sampling interval in our real-world dataset is one minute.
Due to the performance consideration, finer monitoring resolution for each metric is uncommon in OSS.
Thus, different time series will be pre-processed for the same set of timestamps with the same interval of one minute.

% For each fault, $t_{d}$ denotes when the fault is detected \lzy{What is $d$?}.
For each fault, let $t_{d}$ be the \textbf{t}ime a fault is \textbf{d}etected.
We assume that RCA is invoked at $t_{d} + t_{delay}$ to collect necessary information, while it takes data in the period $[t_{d} - t_{ref}, t_{d} + t_{delay}]$ for reference.
The data in $(t_{d} + t_{delay} - t_{test}, t_{d} + t_{delay}]$ are treated as from $P_{\mathbf{m}}$ while $[t_{d} - t_{ref}, t_{d} - t_{test}]$ are taken as fault-free.
By default, we use $t_{delay} = 5$ min, $t_{ref} = 120$ min, and $t_{test} = 10$ min in the experiments.
The effects of different $t_{delay}$ and $t_{ref}$ will be explored in \secref{sec:hyperparameter} with the real-world dataset.
In the rest of this section, we name a fault with its corresponding data in $[t_{d} - t_{ref}, t_{d} + t_{delay}]$ as a \textit{case}.

\subsubsection{Evaluation Metrics}

%We evaluate the performance of a method through the probability that the top-k results recommended by the given method include the root cause metrics, denoted as $AC@k$.
Following existing works~\cite{Wang:2018,Meng:2020,Yu:2021}, we evaluate the performance of a method through the recall with the top-k results, denoted as $AC@k$.
\eqnref{eqn:exp:ac-k} shows the definition of $AC@k$, where $\mathcal{F}$ is a set of faults and $R_{i}(\mathbf{M})$ is the $i$-th result recommended by the method for each fault $\mathbf{M}$.
\eqnref{eqn:exp:ac-k} is slightly different from the evaluation metrics in the previous works~\cite{Wang:2018,Meng:2020,Yu:2021}, ensuring that $AC@k$ is monotonically non-decreasing with $k$.
73\% of developers only consider the top-5 results of a fault localization technique, according to the survey in \cite{Kochhar:2016}.
As a result, we present $AC@k$ for $k \le K = 5$.
Moreover, we show the overall performance by $Avg@K = \frac{1}{K}\sum_{k=1}^{K} AC@k$.
In terms of efficiency, we record analysis duration per fault, denoted as $T$ in the unit of seconds.

\begin{equation}\label{eqn:exp:ac-k}
	AC@k = \frac{1}{|\mathcal{F}|}\sum_{\mathbf{M} \in \mathcal{F}} \frac{|\mathbf{M} \cap \{R_{i}(\mathbf{M}) \mid i = 1, 2, \cdots, k\} |}{\lvert\mathbf{M}\rvert}
\end{equation}

\subsubsection{Baselines}

Each baseline is separated into two steps, namely \textit{graph construction} and \textit{scoring}.
Monitoring metrics will be ranked based on the scores calculated in the final step.
We classify the scoring step in the recent RCA literature for OSS into three groups: DFS-based, random walk-based, and invariant network-based.
In each group, we choose the representative works.
Moreover, we choose the graph construction methods adopted in those works as the baseline ones for the first step.
% We choose baselines for both steps in the recent RCA literature for OSS.

In the graph construction step, the PC algorithm~\cite{Kalisch:2012} is widely used~\cite{Chen:2014,Lin:2018,Wang:2018,Ma:2020}.
We choose Fisher's z-transformation of the partial correlation and $G^{2}$ test as the conditional independence tests for PC, denoted as \textit{PC-gauss} and \textit{PC-gsq}, respectively.
PCMCI~\cite{Runge:2019} adapts PC for time series, based on which \textit{PCTS}~\cite{Meng:2020} transfers the lagged graph into the one among monitoring metrics.
Moreover, the structural graph proposed in this work is denoted as \textit{Structural}.

As for the scoring step, \textit{DFS} traverses the abnormal nodes in the graph, ranking the roots of the sub-graph via anomaly scores~\cite{Chen:2014}.
Its variant \textit{DFS-MS} further ranks candidate metrics according to correlation with the SLI~\cite{Lin:2018}.
Another variant \textit{DFS-MH} traverses the abnormal sub-graph until a node is not correlated with its parents~\cite{Liu:2021}.
The DFS-based methods take the result of anomaly detection as input.
We choose z-score used in \cite{Lin:2018} and SPOT~\cite{Siffer:2017} used in \cite{Meng:2020} as options.
These anomaly detection methods are also taken as baselines\footnote{Anomaly detection and invariant network-based methods will utilize an empty graph with all the available monitoring metrics but no edges.}, denoted as \textit{NSigma} and \textit{SPOT}, respectively.
Another line of works is random walk-based methods.
\textit{RW-Par} calculates the transition probability via partial correlation~\cite{Meng:2020}, while \textit{RW-2} is short for the second-order random walk with Pearson correlation~\cite{Wang:2018}.
\textit{ENMF}\footnote{We take ``ENMF'' from their code to prevent abbreviation duplication between Ranking Causal Anomalies~\cite{Cheng:2016} and Root Cause Analysis.} constructs an invariant network based on the ARX model, explicitly modeling the fault propagation~\cite{Cheng:2016}.
% \footnotemark[ \the\numexpr\value{footnote} - 1\relax]
\textit{CRD} further extends \textit{ENMF} with broken cluster identification~\cite{Ni:2017}.
% \footnote{As the authors' implementation is not available, we implement CRD based on our understanding}

\subsection{Simulation Study}

Three datasets are generated with 50 / 100 / 500 nodes and 100 / 500 / 5,000 edges, respectively, denoted as $\mathcal{D}_{Sim}^{N}$ where $N$ is the number of nodes.
For each dataset, we generate 10 graphs and 100 cases per graph.
Evaluation metrics averaged among the 10 graphs will be presented.
The parameters of baseline methods are selected to achieve the best AC@5 on the first graph in $\mathcal{D}_{Sim}^{50}$.

\subsubsection{Data Generation}

We generate time series based on the Vector Auto-regression model, as shown in \eqnref{eqn:var}.
$\mathbf{x}^{(t)}$ is a column vector of the metrics at time $t$.
$\mathbf{A}$ is the weighted adjacent matrix encoding the CBN.
% $A_{ij} \neq 0$ means the $j$-th metric is a cause of the $i$-th one, where $A_{ij}$ is uniformly sampled from $(-2.0, -0.5) \cup (0.5, 2.0)$ \lzy{why sampling from this range?}, representing the causal effect, \textit{e.g.}, memory usage per request.
$A_{ij} \neq 0$ means the $j$-th metric is a cause of the $i$-th one, where $A_{ij}$ represents the causal effect, \textit{e.g.}, the memory usage per request.
The CBN is enforced to be a connected DAG with only the first node (SLI) having no children.
The item $\beta \mathbf{x}^{(t-1)}$ reflects the auto-regression nature of the time series.
The final item $\mathbf{\epsilon}^{(t)}$ is Gaussian noises, representing the natural fluctuation due to unobserved variables.

\begin{equation}\label{eqn:var}
	\mathbf{x}^{(t)} = \mathbf{A} \mathbf{x}^{(t)} + \beta \mathbf{x}^{(t-1)} + \mathbf{\epsilon}^{(t)}
\end{equation}

To inject a fault $\mathbf{M}$ at time $t$, we first generate the number of root cause metrics $|\mathbf{M}|$.
$|\mathbf{M}| - 1$ follows a Poisson distribution, as it is rare for a fault to affect many metrics directly.
For each $V_{i} \in \mathbf{M}$, the noise item will be altered as $u_{i}^{(t)} = \epsilon_{i}^{(t)} + a_{i} \sigma_{i}$ for 2 timestamps.
The random parameter $a_{i}$ will make the SLI metric abnormal according to the three-sigma rule of thumb.

\subsubsection{Performance Evaluation}

\tableref{tab:performance:sim} summarizes the performance of different methods in three simulation datasets.
The scoring step of each method uses the graph deduced by $\mathbf{A}$ directly, \textit{i.e.}, $X_{j} \in \mathbf{Pa}(X_{i}) \Leftrightarrow \mathbf{A}_{ij} \neq 0$.
We choose the linear regression for RHT.
Moreover, RHT could achieve the best performance in theory if it regards the parents as $\mathbf{Pa}(X_{i}^{(t)}) = \mathbf{Pa}^{(t)}(X_{i}) \cup \left\{X_{i}^{(t-1)}\right\}$.
Such implementation is denoted as \textit{RHT-PG}, where \textit{PG} represents the \textit{perfect graph}.
As the linear relation with the perfect graph performs as the best proxy of $\mathcal{L}_{1}$, we do not consider the descendant adjustment in the simulation study.

RHT-PG approaches the ideal performance, outperforming baseline methods ($p < 0.001$ in t-test for AC@k), which shows the theoretical reliability of our method.
There is a gap between the performance of RHT and RHT-PG, which enlarges as the number of nodes increases.
This phenomenon illustrates the restriction of Corollary~\ref{thm:l2-necessary} that a broken CBN cannot guarantee a correct answer to RCA.
On the other hand, RHT-PG is not perfect yet, which may be the result of statistical errors introduced in hypothesis testing with limited faulty data.

\begin{table*}[tb]
	\caption{
	    Performance of different methods in the simulation study. We put the standard deviation in the parentheses behind each evaluation metric.
	    \textit{RHT-PG} represents \textit{RHT} with the \textit{perfect graph}.
	}\label{tab:performance:sim}
	\begin{tabular}{l ccr ccr ccr}
		\toprule
		\multirow{2}{0.15\columnwidth}{Scoring Method} & \multicolumn{3}{c}{$\mathcal{D}_{Sim}^{50}$} & \multicolumn{3}{c}{$\mathcal{D}_{Sim}^{100}$} & \multicolumn{3}{c}{$\mathcal{D}_{Sim}^{500}$} \\
		& AC@1 & AC@5 & \multirow{1}{0.12\columnwidth}{T (s)} & AC@1 & AC@5 & \multirow{1}{0.12\columnwidth}{T (s)} & AC@1 & AC@5 & \multirow{1}{0.12\columnwidth}{T (s)} \\ 
		\midrule
		NSigma & 0.432(0.05) & 0.733(0.03) & 0.306(0.00) & 0.384(0.05) & 0.613(0.03) & 0.575(0.01) & 0.376(0.04) & 0.579(0.03) & 2.759(0.03) \\
		SPOT & 0.508(0.04) & 0.761(0.03) & 6.601(0.21) & 0.451(0.04) & 0.670(0.03) & 17.365(1.14) & 0.225(0.07) & 0.509(0.07) & 83.465(10.85) \\
		DFS & 0.541(0.04) & 0.682(0.05) & 0.308(0.00) & 0.555(0.03) & 0.653(0.03) & 0.579(0.01) & 0.540(0.03) & 0.611(0.02) & 2.790(0.06) \\
		DFS-MS & 0.515(0.03) & 0.682(0.05) & 0.502(0.00) & 0.517(0.03) & 0.652(0.03) & 0.964(0.01) & 0.191(0.09) & 0.542(0.06) & 4.665(0.06) \\
		DFS-MH & 0.178(0.08) & 0.217(0.09) & 0.501(0.00) & 0.272(0.05) & 0.365(0.05) & 0.969(0.01) & 0.489(0.04) & 0.605(0.03) & 4.735(0.05) \\
		RW-Par & 0.188(0.06) & 0.433(0.07) & 0.714(0.00) & 0.136(0.05) & 0.295(0.07) & 1.761(0.01) & 0.004(0.01) & 0.017(0.02) & 20.246(0.10) \\
		RW-2\footnotemark & 0.188(0.06) & 0.433(0.07) & 0.437(0.00) & 0.136(0.05) & 0.295(0.07) & 1.059(0.01) & 0.004(0.01) & 0.017(0.02) & 10.141(0.11) \\
		ENMF & 0.116(0.03) & 0.278(0.04) & 0.624(0.01) & 0.200(0.03) & 0.336(0.05) & 1.865(0.03) & 0.217(0.04) & 0.354(0.07) & 34.082(0.55) \\
		CRD & 0.074(0.02) & 0.223(0.04) & 4.844(0.03) & 0.013(0.01) & 0.064(0.02) & 6.767(0.10) & 0.003(0.01) & 0.011(0.01) & 46.933(0.74) \\
		RHT & 0.598(0.03) & 0.880(0.02) & 0.338(0.01) & 0.535(0.06) & 0.749(0.06) & 0.658(0.01) & 0.510(0.04) & 0.644(0.04) & 3.326(0.06) \\
		RHT-PG & \textbf{0.615}(0.02) & \textbf{0.952}(0.01) & 0.346(0.00) & \textbf{0.631}(0.02) & \textbf{0.930}(0.01) & 0.665(0.01) & \textbf{0.623}(0.03) & \textbf{0.823}(0.03) & 3.310(0.07) \\
		\midrule
        Ideal & 0.617(0.02) & 0.999(0.00) & & 0.633(0.02) & 0.999(0.00) & & 0.634(0.04) & 1.000(0.00) & \\ 
		\bottomrule
	\end{tabular}
\end{table*}

\footnotetext{RW-2 is degraded to the first-order random walk with its best parameter, hence having the identical performance to RW-Par.\label{footnote:RW-2}}

\subsubsection{Robustness Evaluation}

Faults with the same strength may have different effects on the SLI.
\citeauthor{Yang:2021} name such a phenomenon as the \textit{dependency intensity} in cloud systems, \textit{i.e.}, ``how much the status of the callee service influences the caller service''~\cite{Yang:2021}.
In this simulation study, we further classify faults into three types based on their dependency intensities with the SLI.
We evaluate the performance of RCA methods against faults of each type separately.

\eqnref{eqn:var} can be transformed into $\mathbf{x}^{(t)} = \mathbf{W} (\beta \mathbf{x}^{(t-1)} + \mathbf{\epsilon}^{(t)})$, where $\mathbf{W} = (I - \mathbf{A})^{-1}$.
Notice that $\mathbf{W}$ is well-defined as $\mathbf{A}$ is generated to be a DAG, which does not have full rank.
The element of $\mathbf{W}$ means that $x_{i}$ will increase by $\mathbf{W}_{ij}$ when $x_{j}$ increases by 1.
Denote the standard deviation of $X_{i}$ based on data before fault as $\hat{\sigma}_{i}$.
We classify each fault $\mathbf{M}$ in the simulated datasets into three types:
\begin{description}
	\item[Weak] The root cause metrics deviate from the normal status dramatically to make a slight fluctuation in the SLI (the first node), \textit{i.e.}, $\left(\forall X_{i} \in \mathbf{M}\right) \mathbf{W}_{1i} \hat{\sigma}_{i} / \hat{\sigma}_{1} < 1$;
	\item[Strong] A slight fluctuation in the root cause metrics can change the SLI dramatically, \textit{i.e.}, $\left(\forall X_{i} \in \mathbf{M}\right) \mathbf{W}_{1i} \hat{\sigma}_{i} / \hat{\sigma}_{1} > 1$;
	\item[Mixed] A fault contains metrics with both the above two types or $X_{i}$ with $\mathbf{W}_{1i} \hat{\sigma}_{i} / \hat{\sigma}_{1} = 1$.
\end{description}

\tableref{tab:robustness:sim50} shows the results on $\mathcal{D}_{Sim}^{50}$.
The results on $\mathcal{D}_{Sim}^{100}$ and $\mathcal{D}_{Sim}^{500}$ are omitted since there are only 4 and 5 strong faults in these two datasets, respectively.
RHT and RHT-PG achieve the best results no matter the type of faults, implying that RHT is more robust than baseline methods.
Anomaly detection methods have competitive performance with weak faults.
Their performance drops in strong faults because root cause metrics may be less abnormal than others.
DFS-based methods are sensitive to the results of anomaly detection.
Their performance shares a similar trend with anomaly detection methods, from weak faults to strong ones.

\begin{table}[tb]
	\caption{Robustness evaluation on $\mathcal{D}_{Sim}^{50}$.
	Faults are classified into three types based on their indicators' influence on SLI.}\label{tab:robustness:sim50}
	\setlength{\tabcolsep}{4pt}
	\begin{tabular}{l rr rr rr}
		\toprule
		\multirow{2}{0.12\columnwidth}{Scoring Method} & \multicolumn{2}{c}{Weak (n=916)} & \multicolumn{2}{c}{Mixed (n=64)} & \multicolumn{2}{c}{Strong (n=20)} \\
		& AC@1 & AC@5 & AC@1 & AC@5 & AC@1 & AC@5 \\ 
		\midrule
		NSigma & 0.454 & 0.753 & 0.249 & 0.498 & 0.000 & 0.550 \\
		SPOT & 0.534 & 0.783 & 0.293 & 0.503 & 0.000 & 0.550 \\
		DFS & 0.558 & 0.707 & 0.282 & 0.368 & 0.550 & 0.550 \\
		DFS-MS & 0.531 & 0.707 & 0.277 & 0.368 & 0.550 & 0.550 \\
		DFS-MH & 0.184 & 0.223 & 0.069 & 0.123 & 0.250 & 0.250 \\
		RW-Par & 0.194 & 0.445 & 0.142 & 0.300 & 0.050 & 0.300 \\
		RW-2\textsuperscript{\ref{footnote:RW-2}} & 0.194 & 0.445 & 0.142 & 0.300 & 0.050 & 0.300 \\
		ENMF & 0.111 & 0.269 & 0.124 & 0.321 & 0.300 & 0.550 \\
		CRD & 0.071 & 0.207 & 0.088 & 0.353 & 0.150 & 0.550 \\
		RHT & 0.613 & 0.888 & 0.325 & 0.730 & 0.800 & 1.000 \\
        RHT-PG & \textbf{0.624} & \textbf{0.954} & \textbf{0.358} & \textbf{0.914} & \textbf{1.000} & \textbf{1.000} \\
		\midrule
        Ideal & 0.627 & 1.000 & 0.358 & 0.995 & 1.000 & 1.000 \\ 
		\bottomrule
	\end{tabular}
\end{table}

\subsection{Empirical Study on Oracle Database Data}

We further evaluate different methods in a real-world dataset, denoted as $\mathcal{D}_{O}$.
There are 99 cases in $\mathcal{D}_{O}$.
Each case comes from Oracle databases with high AAS faults in a large banking system.
We choose the parameters of baseline methods for better AC@5.
% We choose the parameters, including the graph construction method, for each scoring method for better AC@5.

\subsubsection{Implementation}

We manually extract the call graph in an Oracle database instance from the official documentation\footnote{Oracle Database Concepts. \url{https://docs.oracle.com/cd/E11882_01/server.112/e40540/}}.
After that, we map 197 monitoring metrics to meta metrics in the skeleton.
The final structural graph contains 2,641 edges.
Oracle database instances may have different sets of metrics.
Therefore, we construct the structural graph for each instance with monitored metrics.

In this empirical study, the ground truth graph is unavailable.
Hence, we compare graph construction methods for each scoring method, choosing the graph with the highest AC@5.
Meanwhile, there is no perfect proxy of $\mathcal{L}_{1}$ (like the CBN and linear relation in the simulation study).
As a result, we fail to include the ideal implementation of RHT (RHT-PG) in the experiment.
We choose the Support Vector Regression (SVR) as the regression method for RHT, which will be discussed in Appendix~\ref{sec:regression-method-selection}.
To alleviate the bias in hypothesis testing, we equip RHT with the descendant adjustment, denoted as \textit{CIRCA}.

\subsubsection{Performance Evaluation}

CIRCA achieves the best results compared with baseline methods, as shown in \tableref{tab:performance:oracle}.
% \lzy{How the combinations of scoring and graph construction methods are selected in this table?}
Random walk-based methods achieve their best performance with PCTS while taking much time to construct the graph.
With the structural graph, DFS-based methods and CIRCA recommend root cause metrics within seconds.
% Anomaly detection alone has competitive performance, which is consistent with the simulation study.

\begin{table}[tb]
	\caption{Performance of different methods on $\mathcal{D}_{O}$}\label{tab:performance:oracle}
	\begin{tabular}{ll rrrr}
		\toprule
		\multirow{2}{0.12\columnwidth}{Scoring Method} & \multirow{2}{0.12\columnwidth}{Graph Method} & \multirow{2}{*}{AC@1} & \multirow{2}{*}{AC@5} & \multirow{2}{*}{Avg@5} & \multirow{2}{0.08\columnwidth}{T (s)} \\
		 & & & & & \\
		\midrule
		NSigma & Empty & 0.323 & 0.662 & 0.525 & 0.472 \\
		SPOT & Empty & 0.152 & 0.419 & 0.296 & 5.027 \\
		DFS & Structural & 0.187 & 0.313 & 0.271 & 0.483 \\
		DFS-MS & Structural & 0.207 & 0.308 & 0.275 & 0.839 \\
		DFS-MH & Structural & 0.268 & 0.439 & 0.372 & 0.844 \\
		RW-Par & PCTS & 0.086 & 0.449 & 0.290 & 24.695 \\
		RW-2\textsuperscript{\ref{footnote:RW-2}} & PCTS & 0.086 & 0.449 & 0.290 & 24.559 \\
		ENMF & Empty & 0.111 & 0.374 & 0.254 & 0.771 \\
		CRD & Empty & 0.035 & 0.313 & 0.165 & 4.787 \\
		CIRCA & Structural & \textbf{0.404} & \textbf{0.763} & \textbf{0.603} & 0.578 \\
		\midrule
        \multicolumn{2}{c}{Ideal} & 0.929 & 1.000 & 0.986 & \\ 
		\bottomrule
	\end{tabular}
\end{table}

\begin{figure}[tb]
	\includegraphics[width=0.83\columnwidth]{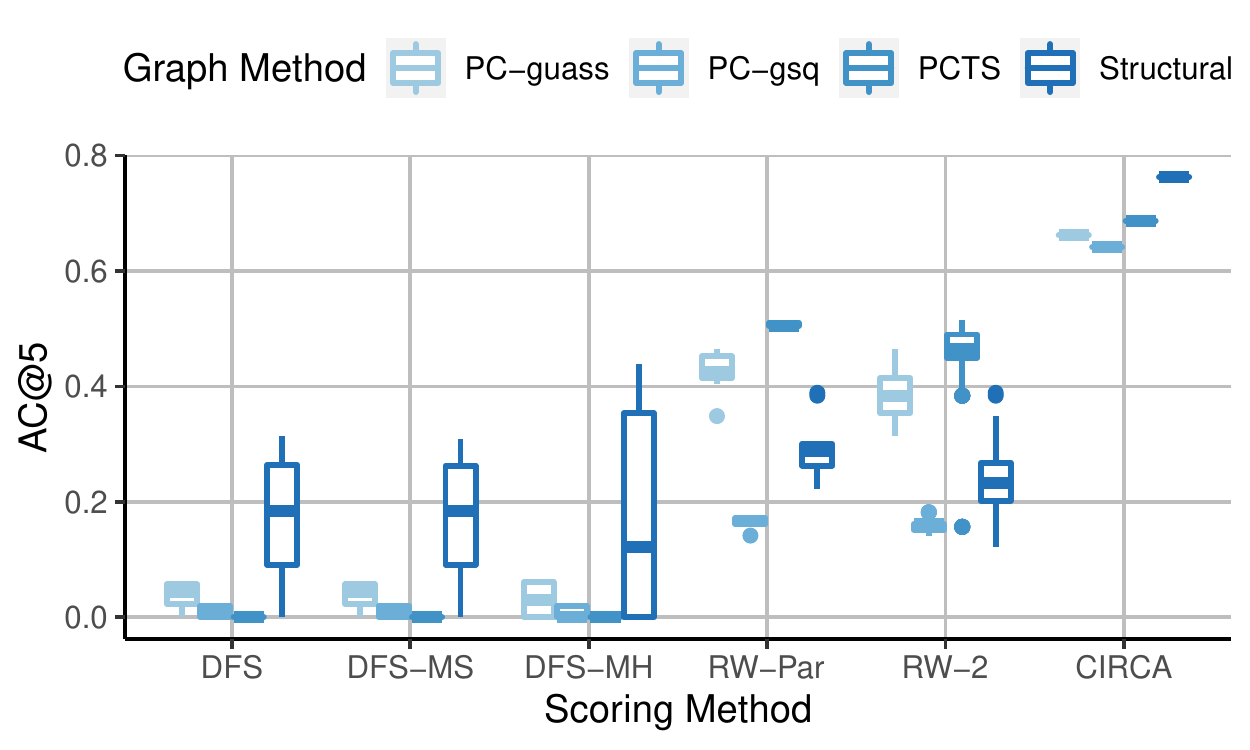}
	\caption{AC@5 for different combinations of ranking methods and graph construction ones}\label{fig:performance:graph-contribution}
	\Description{
		This is a box plot for the combination of ranking methods and graph construction methods.
		The boxes for DFS-based methods with PC-gsq or PCTS have AC@5 of zero, while the boxes for DFS-based methods with the proposed structural graph are higher than those with PC-gauss.
		As for random walk-based ranking methods, the boxes with PCTS are higher than other graph construction methods.
	}
\end{figure}

We remove components from CIRCA progressively to show their contribution, summarized in \tableref{tab:performance:oracle:components}.
The result illustrates that both regression-based hypothesis testing and descendant adjustment have a positive effect.
\figref{fig:performance:graph-contribution} compares the proposed structural graph with other graph construction baselines.
We exclude anomaly detection and invariant network-based methods from this figure, as they cannot utilize the CBN.
% \lzy{Why only part of scoring methods are presented in this figure?}
Each box in \figref{fig:performance:graph-contribution} presents the distribution of AC@5 for a scoring method with different parameters.
One data point is the best AC@5 from different graph construction parameters with the same scoring ones.
The 3 horizontal lines of each box show 25th, 50th, and 75th percentile, while two whiskers extend to minimum and maximum.
The proposed structural graph improves AC@5 for DFS-based methods and CIRCA, while PCTS fits random walk-based methods better.

\begin{table}[tb]
	\caption{
	Contribution of CIRCA's components on $\mathcal{D}_{O}$ with the structural graph.
	}\label{tab:performance:oracle:components}
	\begin{tabular}{lrrrrr}
		\toprule
		\multirow{2}{0.12\columnwidth}{Scoring Method} & \multirow{2}{*}{AC@1} & \multirow{2}{*}{AC@3} & \multirow{2}{*}{AC@5} & \multirow{2}{*}{Avg@5} & \multirow{2}{0.08\columnwidth}{T (s)} \\
		& & & & & \\
		\midrule
		NSigma & 0.323 & 0.586 & 0.662 & 0.525 &  0.472  \\
		RHT & 0.328 & 0.601 & 0.677 & 0.546 & 0.576 \\
		CIRCA & \textbf{0.404} & \textbf{0.616} & \textbf{0.763} & \textbf{0.603} & 0.578 \\
		\bottomrule
	\end{tabular}
\end{table}

\subsubsection{Case Study}

\begin{figure}[tb]
	\includegraphics{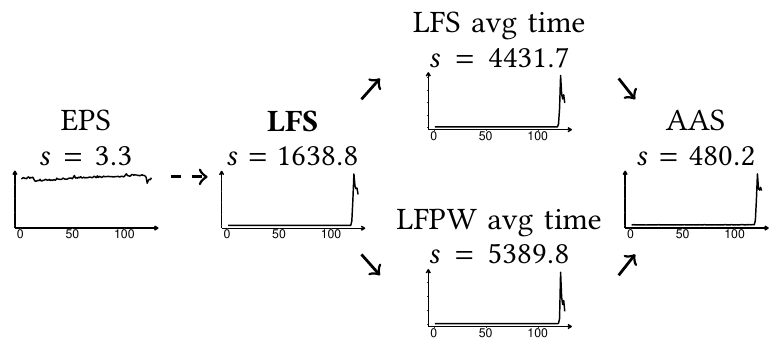}
	\caption{
	    Part of an Oracle database failure, where LFS is the root cause metric labeled by the DBAs.
	   % Below each metric name is the anomaly score calculated by \eqnref{eqn:anomaly-score} and the time series at the same period.
	    Below each metric name is the score calculated by \eqnref{eqn:anomaly-score} and the time series at the same period.
	    Time (horizontal axis) is shown in minutes.
	   % We hide the value (vertical axis) due to the confidentiality policy of our partner company.
	}\label{fig:performance:case}
	\Description{
	    This figure shows part of an Oracle database failure.
	    LFS is the root cause metric.
	    One of its parent, EPS, has an anomaly score slightly larger than 3.
	    On the other hand, the anomaly scores of LFS and its descendants are higher than 480.
	}
\end{figure}

\figref{fig:performance:case} presents a failure, where ``log file sync'' (LFS) is the root cause metric labeled by the database administrators (DBAs).
A poor understanding of $\mathcal{L}_{1}$ puzzles RCA methods.
On the one hand, DFS fails to stop at LFS and continues to check ``execution per second'' (EPS), missing the desired answer.
No baseline method recommends LFS in the top-5 results, except NSigma, ENMF, and CRD.
% \lzy{How does other baselines perform on this cases? For example, RW?}
On the other hand, CIRCA assigns a high anomaly score for AAS after revising it from 532.4 (given by NSigma) to 480.2 with regression.

% CIRCA scores each metric separately to prevent missing the desired answer \lzy{I did not get how it helps}.
DFS-based methods will drop descendants once meeting an abnormal metric.
In contrast, CIRCA scores each metric separately, preventing missing answers like DFS-based methods.
Moreover, CIRCA adjusts the anomaly score of LFS with that of the average time of ``log file parallel write'' (LFPW), \textit{i.e.}, $s_{LFS}^{\prime} = 7028.6$.
% showing the preference for the former \lzy{Maybe add more description about the anomaly score adjustion}.
This technique helps CIRCA rank LFS ahead of the other metrics.

\subsubsection{Lessons Learned}

CIRCA outperforms baseline methods on $\mathcal{D}_{O}$, consistent with the simulation study.
\tableref{tab:performance:oracle:components} and \figref{fig:performance:graph-contribution} further illustrate that each of the 3 proposed techniques has a positive effect.

Though RCA is a difficult task related to Layer 2 of the causal ladder (Corollary~\ref{thm:l2-necessary}), the knowledge of Layer 1 ($\mathcal{L}_{1}$) is incomplete.
We illustrate the negative effect through a case study.
We believe that further advancement in the future has to handle this obstacle explicitly.
At present, we prefer CIRCA to pure RHT if deployed.
Meanwhile, the effectiveness of the descendant adjustment has to be verified on more real-world datasets.

% \subsection{Case Study on Bank Information System Data}

% \textbf{TODO:}
% \cite{Cheng:2016} evaluates their method on a Bank Information System (BIS) dataset, which is in their repository.

\subsection{Discussion}

\subsubsection{Hyperparameter Sensitivity}\label{sec:hyperparameter}

\begin{figure}[tb]
	\includegraphics[width=0.9\columnwidth]{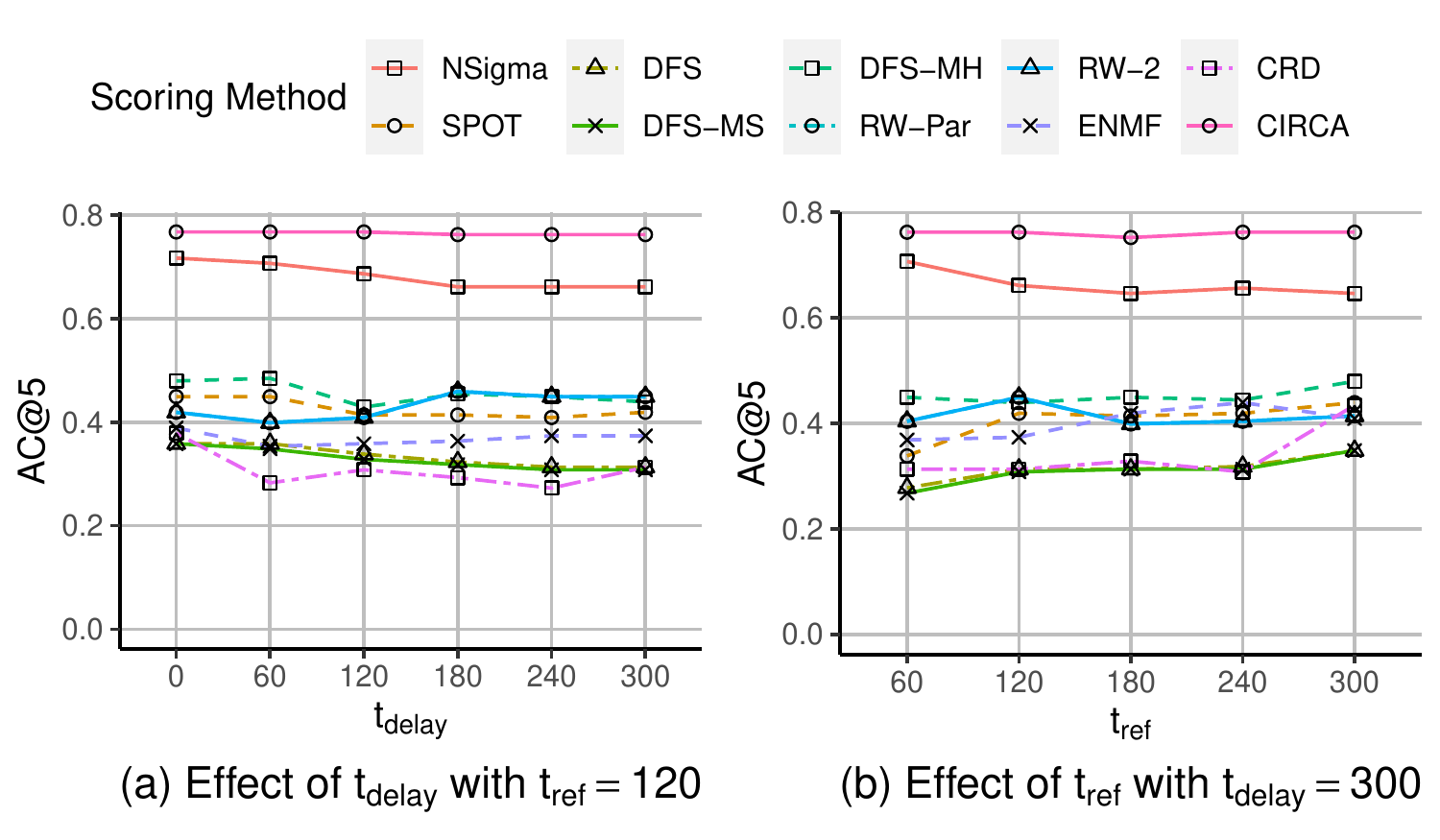}
	\caption{Performance with various hyperparameters on $\mathcal{D}_{O}$}\label{fig:param-sensitivity}
	\Description{
		There are two line charts, showing the performance of baseline methods with different hyperparameters.
		The left one fixes $t_{ref} = 120$, varying $t_delay$, while the right one fixes $t_{delay} = 300$, varying $t_ref$.
	}
\end{figure}

% \figref{fig:param-sensitivity} shows the performance of each method with different $t_{delay}$ and $t_{ref}$.
\figref{fig:param-sensitivity} compares RCA methods with different $t_{delay}$ and $t_{ref}$.
CIRCA has stable performance with these two hyperparameters, outperforming baseline methods.

\subsubsection{Performance of Existing Methods}

% DFS-based methods are sensitive to anomaly detection.
% If we cannot reach the SLI from the root cause metrics along with abnormal nodes, a DFS-based method cannot locate the root cause.

RW-Par and RW-2 represent the scoring methods of MicroCause~\cite{Meng:2020} and CloudRanger~\cite{Wang:2018}, respectively.
However, RW-Par (RW-2) fails to achieve the performance in the corresponding paper.
MicroCause utilizes metric priority provided by operators, which is unavailable for RW-Par.
On the other hand, CloudRanger achieves its best result with a sampling interval of 5 seconds.
% Though it is feasible with trace data, it is impractical for massive metrics.
The coarse monitoring frequency of $\mathcal{D}_{O}$ may explain the poor performance of RW-2.

As stated by Corollary~\ref{thm:l2-necessary}, knowledge of Layer 2 (such as $\mathbf{Pa}$) is necessary for RCA.
The invariant network-based methods utilize the observation data only (Layer 1).
Their unsatisfying performance illustrates the restriction of CHT~\cite{Bareinboim:2022}.

\subsubsection{Feasibility}

\noindent
\textbf{Graph Construction.}
% \paragraph{Graph Construction}
The construction of the proposed structural graph requires system architecture and a mapping from monitoring metrics to the targets to be monitored.
The former is usually in the form of documentation.
We argue that a metric is neither insightful nor actionable unless operators understand its underlying meaning.
Operators need to classify each distinct metrics only once to obtain the mapping.
The mapping can be shared among similar instances of the same type (like Oracle database instances).

\noindent
\textbf{Scalability.}
% \paragraph{Scalability}
As shown in \tableref{tab:performance:sim}, RHT's time cost grows around linearly with the size of the dataset.
% \lzy{It seems CIRCA costs too much time on a large-scale system (tens of thousands of instances)}
Moreover, the design of CIRCA supports horizontal scalability to handle large-scale systems via adding computing resources, as each metric is scored separately.
We plan to train the regression models offline to speed up online analysis.
Mature parallel programming frameworks, such as Apache Spark, may further help accelerate CIRCA.

% !TEX root=../main.tex
\section{Related Works}\label{sec:related-work}

\noindent
\textbf{Root Cause Analysis.}
% \paragraph{Root Cause Analysis}
Corollary~\ref{thm:l2-necessary} explains that graph construction is a common step in the RCA literature for online service system operation.
DFS-based methods~\cite{Chen:2014,Lin:2018,Liu:2021} traverse abnormal sub-graph, which is sensitive to anomaly detection results.
Some works adopt random walk~\cite{Wang:2018,Meng:2020,Ma:2020} or PageRank~\cite{Wang:2021} to score candidate root cause indicators, lacking explainability.
% Random walk-based and PageRank-based methods~\cite{Wang:2018,Meng:2020,Ma:2020,Wang:2021} lack explainability.
Another line of works is invariant network-based methods~\cite{Cheng:2016,Ni:2017}.
As these works adopt the pair-wise manner to learn the invariant relations, it is hard for them to reach the knowledge of RCA, restricted by CHT~\cite{Bareinboim:2022}.
No methods above utilize causal inference.
Sage~\cite{Gan:2021} conducts counterfactual analysis to locate root causes without a formal formulation.
Corollary~\ref{thm:l3-unnecessary} states that counterfactual analysis is unnecessary.
Hence, we did not include this method as a baseline.
Meanwhile, CHT~\cite{Bareinboim:2022} indicates that it can be hard to conduct counterfactual analysis even with a CBN.

The definition of root cause analysis varies with the scenario in the literature.
% We treat the observed projection of a fault as the desired answer.
% In contrast, some applications require an answer beyond the data, taking RCA as a classification task with supervised learning~\cite{Yi:2021}.
% For homogeneous devices or services, operators are interested in the common features~\cite{Zhang:2021}.
% Accordingly, multi-dimensional root cause analysis is conducted.
Some applications require an answer beyond the data, taking RCA as a classification task with supervised learning~\cite{Yi:2021}.
For homogeneous devices or services, operators are interested in the common features~\cite{Zhang:2021}.
Accordingly, a multi-dimensional root cause analysis is conducted.
In contrast, we treat the observed projection of a fault as the desired answer.
The \nameref{thm:rc-criterion} further relates RCA in this work with \textit{contextual anomaly detection}~\cite{Chandola:2009}, treating parents in the CBN as the context for each variable.
We take complex contextual anomaly detection methods as future work.

\noindent
\textbf{Causal Discovery.}
% \paragraph{Causal Discovery}
The task to obtain the CBN is named \textit{causal discovery}.
We refer the readers to a recent survey~\cite{Guo:2020} for a thorough discussion.
NOTEARS~\cite{Zheng:2018} converts the DAG search problem from the discrete space into a continuous one.
Following NOTEARS, some recent works are based on gradient descent~\cite{He:2021}.

Although causal discovery has its sound theory, the CBN discovered from data directly is not explainable for human operators.
In contrast, some works obtain the CBN based on domain knowledge.
MicroHECL~\cite{Liu:2021} traces the fault along with traffic, latency, or error rate in the call graph.
Meanwhile, Sage~\cite{Gan:2021} constructs the CBN among latency and machine metrics.
The structural graph proposed in this work is compatible with the assumptions in these two works, extending the kinds of meta metrics.

% !TEX root=../main.tex
\section{Conclusion and Future Work}\label{sec:conclusion}

Root cause analysis (RCA) is an essential task for OSS operations.
In this work, we formulate RCA as a new causal inference task named \textit{intervention recognition}, based on which, we further obtain the \nameref{thm:rc-criterion} to find the root cause.
We believe such a formulation bridge two well-studied fields (RCA and causal inference) and provide a promising new direction for the critical-yet-hard-to-solve RCA problem in OSS.

% To apply such a criterion in OSS, we propose a novel unsupervised causal inference-based RCA method, \textit{CIRCA}.
To apply such a criterion in OSS, we propose a novel causal inference-based RCA method, \textit{CIRCA}.
CIRCA consists of three techniques, namely structural graph construction, regression-based hypothesis testing, and descendant adjustment.
We verify the theoretical reliability of CIRCA in the simulation study.
Moreover, CIRCA also outperforms baseline methods in a real-world dataset.

In the future, we plan to include faulty data for regression.
We hope that diverse data can help overcome the limited understanding of the system's normal status.
This work rests on a set of assumptions that a real application may not satisfy.
For example, some meta metrics do not have corresponding monitoring metrics in the skeleton we construct for the Oracle database.
% In other words, they are unobserved common drivers of the downstream ones, violating the Causal Sufficiency assumption.
As a result, they can imply common exogenous parents of the downstream monitoring metrics, violating the Markovian assumption.
Explicitly modeling these hidden meta metrics may improve RCA performance.
Meanwhile, the retrospect of analysis mistakes may also point to the lack of monitoring.
Beyond the analysis framework in this work, discoveries on the underlying mechanism of OSS can also help climb the ladder of causation for the RCA task.

\begin{acks}
	We thank Li Cao, Zhihan Li, and Yuan Meng for their helpful discussions on this work, thank Xianglin Lu for her data preparation work, and thank Xiangyang Chen, Duogang Wu, and Xin Yang for sharing their knowledge on Oracle databases.
	This work is supported by the \grantsponsor{NKRDPC}{National Key R\&D Program of China}{} under Grant \grantnum{NKRDPC}{2019YFB1802504}, and the \grantsponsor{SKPNNSC}{State Key Program of National Natural Science of China}{} under Grant \grantnum{SKPNNSC}{62072264}.
\end{acks}

\bibliographystyle{ACM-Reference-Format}
\bibliography{bibliography}

\clearpage
\appendix

% !TEX root=../main.tex
\section{Proof of Theorem \ref{thm:rca-l2}}\label{sec:proof-rca-l2}

We define \textit{identifiable intervention recognition} (IIR) as follows:

\begin{definition}[Identifiable Intervention Recognition, IIR]\label{def:identifiable-intervention-recognition}
    % Based on $\mathcal{L}_{1}$ and $P_{\mathbf{m}}$
	Identifiable intervention recognition is to find out a set of potential interventions $\{\mathbf{m}^{\prime} \mid \mathcal{L}_{1}(\mathbf{V} \mid do(\mathbf{m}^{\prime})) \equiv P_{\mathbf{m}} \}$ (\textit{i.e.}, \textit{identifiable interventions}).
\end{definition}

With Definition~\ref{def:identifiable-intervention-recognition}, we have Lemma~\ref{lemma:l2-to-rca} and Lemma~\ref{lemma:rca-to-l2}.

\begin{lemma}\label{lemma:l2-to-rca}
	The knowledge of IIR can be derived from $\mathcal{L}_{2}$.
\end{lemma}
\begin{proof}[Proof of Lemma~\ref{lemma:l2-to-rca}]
	Denote $\mathcal{C}$ as the equivalence classes defined by $\mathcal{L}_{2}$ among $\mathcal{F} = \bigcup_{\mathbf{M} \in 2^{\mathbf{V}}} Val(\mathbf{M})$, where $\mathcal{F}$ is all possible interventions, including no intervention.
	For each equivalent class $c \in \mathcal{C}$, $c$ is a set of interventions $[\mathbf{m}]$ which leads to the same distribution over $\mathbf{V}$:
	$$[\mathbf{m}] = \{\mathbf{m}^{\prime} \mid \mathcal{L}_{1}(\mathbf{V} \mid do(\mathbf{m}^{\prime})) \equiv P_{\mathbf{m}}\}$$
	Denote the distribution of V under $[\mathbf{m}]$ as $\mathcal{L}_{2}^{-1}(P_{\mathbf{m}}) = [\mathbf{m}]$.
	Denote $\mathcal{L}_{2}^{\prime}([\mathbf{m}]) = P_{\mathbf{m}}$, where $[\mathbf{m}] \in \mathcal{C}$.
	For any $\mathbf{m}_{1}, \mathbf{m}_{2} \in \mathcal{F}$, we always have:
	$$P_{\mathbf{m}_{1}} \equiv P_{\mathbf{m}_{2}} \rightarrow [\mathbf{m}_{1}] = [\mathbf{m}_{2}]$$
	Hence, $\mathcal{L}_{2}^{\prime}$ is a one-to-one correspondence and have its inverse mapping, denoted as $\mathcal{L}_{2}^{-1}(P_{\mathbf{m}}) = [\mathbf{m}]$.

	When an intervention occurs, based on $P_{\mathbf{m}} \in \mathcal{L}_{2}(\mathcal{F})$, $\mathcal{L}_{2}^{-1}(P_{\mathbf{m}})$ is the set of targets of IIR.
	Hence, the knowledge of IIR can be derived from $\mathcal{L}_{2}$.
\end{proof}

\begin{lemma}\label{lemma:rca-to-l2}
	The knowledge of IIR encodes $\mathcal{L}_{2}$.
\end{lemma}
\begin{proof}[Proof of Lemma~\ref{lemma:rca-to-l2}]
	For any valid $P_{\mathbf{m}} \in \mathcal{L}_{2}(\mathcal{F})$, IIR will produce a set of possible interventions $\{\mathbf{m}^{\prime} \mid \mathcal{L}_{2}(\mathbf{m}^{\prime}) \equiv P_{\mathbf{m}}\}$.
% 	Notice that the corresponding assignment $\mathbf{m}^{\prime}$ for each possible root cause is just encoded in $P_{\mathbf{m}^{\prime}}$.
	Hence, the knowledge of IIR can be extended to the mapping $\mathcal{L}_{2}^{-1}$ that maps $P_{\mathbf{m}}$ to the element of $\mathcal{C}$.
	Meanwhile, $\mathcal{L}_{2}^{-1}$ is the inverse mapping of $\mathcal{L}_{2}^{\prime}$ and $\mathcal{L}_{2}$ can be derived from $\mathcal{L}_{2}^{\prime}$.
	Hence, the knowledge of IIR encodes $\mathcal{L}_{2}$.
\end{proof}

Based on Lemma~\ref{lemma:l2-to-rca} and Lemma~\ref{lemma:rca-to-l2}, the knowledge of IIR is equivalent to $\mathcal{L}_{2}$.
Then we have \thmref{thm:irca-l2}.

\begin{theorem}\label{thm:irca-l2}
	The knowledge of IIR is at the second layer of the causal ladder.
\end{theorem}

For an SCM with a CBN, we get Lemma~\ref{lemma:decompose-fault}.

\begin{lemma}\label{lemma:decompose-fault}
	For a given SCM $\mathcal{M}$ with a CBN $\mathcal{G}$, let $\mathbf{Pa}(V_{i})$ be the parents of $V_{i}$ in $\mathcal{G}$.
	$P(V_{i} \mid \mathbf{pa}(V_{i}), do(\mathbf{m}))$ can be reduced to the form defined in \eqnref{eqn:parent-block-fault}.
	\begin{equation}\label{eqn:parent-block-fault}
		P(V_{i} \mid \mathbf{pa}(V_{i}), do(\mathbf{m})) = \left\{\begin{array}{ll}
			P(V_{i} \mid \mathbf{pa}(V_{i}), do(v_{i})), & V_{i} \in \mathbf{M} \\
			P(V_{i} \mid \mathbf{pa}(V_{i})), & V_{i} \notin \mathbf{M} \\
		\end{array}\right.
	\end{equation}
\end{lemma}
\begin{proof}[Proof of Lemma~\ref{lemma:decompose-fault}]
	Given a variable $V_{i}$, the intervened variables $\mathbf{M}$ can be separated into three parts: 1) $\mathbf{M}_{Pa} = \mathbf{M} \cap \mathbf{Pa}(V_{i})$, 2) $\mathbf{M}_{i} = \mathbf{M} \cap \{V_{i}\}$, and 3) $\mathbf{M}_{Other} = \mathbf{M} \setminus \mathbf{Pa}(V_{i}) \setminus \{V_{i}\}$.
	Hence, 
	\begin{itemize}
		\item $\mathbf{M} = \mathbf{M}_{Pa} \cup \mathbf{M}_{i} \cup \mathbf{M}_{Other}$,
		\item $\mathbf{M}_{Pa} \cap \mathbf{M}_{i} =  \mathbf{M}_{Pa} \cap \mathbf{M}_{Other} = \mathbf{M}_{i}\cap \mathbf{M}_{Other} = \emptyset$, and
		\item there is no arrow from $\mathbf{M}_{Other}$ to $V_{i}$ in $\mathcal{G}$.
	\end{itemize}

    \paragraph{Case 1.}
    With $V_{i} \in \mathbf{M}$, denote the interventional SCM of $\mathcal{M}$ with $do(\mathbf{m})$ as $\mathcal{M}^{\prime}$.
    $\mathcal{M}^{\prime}$ replaces $f_{i}$ in $\mathcal{M}$ with $v_{i} \gets m_{i}$~\cite{Pearl:2009}.
    As a result, other variables cannot affect  $V_{i}$ any longer.
    Hence, $do(v_{i})$ makes $P(V_{i} \mid \mathbf{pa}(V_{i}), do(\mathbf{m}))$ and $P(V_{i} \mid \mathbf{pa}(V_{i}), do(v_{i}))$ equivalent by $$P(V_{i} \mid \mathbf{pa}(V_{i}), do(\mathbf{m})) = P(V_{i} \mid do(v_{i})) = P(V_{i} \mid \mathbf{pa}(V_{i}), do(v_{i}))$$

    \paragraph{Case 2.}
    With $V_{i} \notin \mathbf{M}$, we get $\mathbf{M}_{i} = \emptyset$ and $\mathbf{M} = \mathbf{M}_{Pa} \cup \mathbf{M}_{Other}$ for the equation below
    $$P(V_{i} \mid \mathbf{pa}(V_{i}), do(\mathbf{m})) = P(V_{i} \mid \mathbf{pa}(V_{i}), do(\mathbf{m}_{Pa}), do(\mathbf{m}_{Other}))$$
    Since $\mathcal{G}$ is a CBN, the ``Parents do/see'' condition~\cite{Bareinboim:2022} states that we can replace $\mathbf{pa}(V_{i})$ with $do(\mathbf{pa}(V_{i}))$.
	\begin{equation}
		P(V_{i} \mid \mathbf{pa}(V_{i}), do(\mathbf{m})) = P(V_{i} \mid do(\mathbf{pa}(V_{i})), do(\mathbf{m}_{Other}))
	\end{equation}
	As we already take $do(\mathbf{m}_{Pa})$ as the condition, the ``Missing-link'' condition~\cite{Bareinboim:2022} ensures that we can drop $do(\mathbf{m}_{Other})$.
	\begin{equation}
		P(V_{i} \mid \mathbf{pa}(V_{i}), do(\mathbf{m})) = P(V_{i} \mid do(\mathbf{pa}(V_{i})))
	\end{equation}
	With the ``Parents do/see'' condition~\cite{Bareinboim:2022} again,  we obtain
	\begin{equation}
		P(V_{i} \mid \mathbf{pa}(V_{i}), do(\mathbf{m})) = P(V_{i} \mid \mathbf{pa}(V_{i}))
	\end{equation}

	Combining the two cases above provides the final conclusion.
\end{proof}

\begin{proof}[Proof of \thmref{thm:rca-l2}]
	Given an intervention $\mathbf{m}$ and any identifiable intervention $\mathbf{m}^{\prime}$ provided by IIR, $P(\mathbf{V} \mid do(\mathbf{m})) \equiv P(\mathbf{V} \mid do(\mathbf{m}^{\prime}))$.
	Hence, $P(V_{i} \mid \mathbf{pa}(V_{i}), do(\mathbf{m})) \equiv P(V_{i} \mid \mathbf{pa}(V_{i}), do(\mathbf{m}^{\prime}))$ for any $V_{i} \in \mathbf{V}$.
	Notice that the corresponding assignment for an intervention is just encoded in the interventional distribution.
	As a result, $(\forall x \in \mathbf{m}, x^{\prime} \in \mathbf{m}^{\prime}) X=X^{\prime} \rightarrow x=x^{\prime}$.
% 	Notice that the corresponding assignment $\mathbf{m}^{\prime}$ for each possible root cause is just encoded in $P_{\mathbf{m}^{\prime}}$.

	Assume that $\mathbf{m}$ is different from $\mathbf{m}^{\prime}$, \textit{e.g.}, $(\exists X \in \mathbf{V}) X \in \mathbf{M} \wedge X \notin \mathbf{M}^{\prime}$.
	With Lemma~\ref{lemma:decompose-fault}, we have $P(X \mid \mathbf{pa}(X), do(x)) \equiv P(X \mid \mathbf{pa}(X))$, which violates the Faithfulness assumption.
	It is the same for the case $(\exists X \in \mathbf{V}) X \notin \mathbf{M} \wedge X \in \mathbf{M}^{\prime}$.
	Hence, IIR can distinguish $\mathbf{m}$ from other interventions, providing the same answer as IR.

	According to \thmref{thm:irca-l2}, we reach the conclusion that the knowledge of IR is at the second layer of the causal ladder.
\end{proof}

\section{Proof of Theorem \ref{thm:rc-criterion}}\label{sec:proof-rc-criterion}

\begin{proof}[Proof of \thmref{thm:rc-criterion}]
	Notice that $P_{\mathbf{m}}(V_{i} \mid \mathbf{pa}(V_{i})) = P(V_{i} \mid \mathbf{pa}(V_{i}), do(\mathbf{m}))$, while $\mathcal{L}_{1}(V_{i} \mid \mathbf{pa}(V_{i})) = P(V_{i} \mid \mathbf{pa}(V_{i}))$.

    \paragraph{Case 1.}
    With $V_{i} \in \mathbf{M}$, we get $\mathbf{M}_{i} = \{V_{i}\}$.
	Under the Faithfulness assumption, \eqnref{eqn:faithfulness-conditional} must hold, while Lemma~\ref{lemma:decompose-fault} provides $P(V_{i} \mid \mathbf{pa}(V_{i}), do(\mathbf{m})) = P(V_{i} \mid \mathbf{pa}(V_{i}), do(v_{i}))$.
	\begin{equation}\label{eqn:faithfulness-conditional}
		P(V_{i} \mid \mathbf{pa}(V_{i}), do(v_{i})) \neq P(V_{i} \mid \mathbf{pa}(V_{i}))
	\end{equation}
	Hence, $$V_{i} \in \mathbf{M} \Rightarrow P_{\mathbf{m}}(V_{i} \mid \mathbf{pa}(V_{i})) \neq \mathcal{L}_{1}(V_{i} \mid \mathbf{pa}(V_{i}))$$
	
	\paragraph{Case 2.}
	With $V_{i} \notin \mathbf{M}$, Lemma~\ref{lemma:decompose-fault} provides $P(V_{i} \mid \mathbf{pa}(V_{i}), do(\mathbf{m})) = P(V_{i} \mid \mathbf{pa}(V_{i}))$.
	Hence, $$V_{i} \notin \mathbf{M} \Rightarrow P_{\mathbf{m}}(V_{i} \mid \mathbf{pa}(V_{i})) = \mathcal{L}_{1}(V_{i} \mid \mathbf{pa}(V_{i}))$$
	Its contrapositive stands as well, $$P_{\mathbf{m}}(V_{i} \mid \mathbf{pa}(V_{i})) \neq \mathcal{L}_{1}(V_{i} \mid \mathbf{pa}(V_{i})) \Rightarrow V_{i} \in \mathbf{M}$$, which is the converse proposition of Case 1.

% 	\begin{enumerate}[label=Case. \arabic*]
% 		\item With $V_{i} \in \mathbf{M}$, we get $\mathbf{M}_{i} = \{V_{i}\}$.
% 		Under the Faithfulness assumption, \eqnref{eqn:faithfulness-conditional} must hold, while Lemma~\ref{lemma:decompose-fault} provides $P(V_{i} \mid \mathbf{pa}(V_{i}), do(\mathbf{m})) = P(V_{i} \mid \mathbf{pa}(V_{i}), do(v_{i}))$.
% 		\begin{equation}\label{eqn:faithfulness-conditional}
% 			P(V_{i} \mid \mathbf{pa}(V_{i}), do(v_{i})) \neq P(V_{i} \mid \mathbf{pa}(V_{i}))
% 		\end{equation}
% 		Hence, $$V_{i} \in \mathbf{M} \Rightarrow P_{\mathbf{m}}(V_{i} \mid \mathbf{pa}(V_{i})) \neq \mathcal{L}_{1}(V_{i} \mid \mathbf{pa}(V_{i}))$$

% 		\item With $V_{i} \notin \mathbf{M}$, Lemma~\ref{lemma:decompose-fault} provides $P(V_{i} \mid \mathbf{pa}(V_{i}), do(\mathbf{m})) = P(V_{i} \mid \mathbf{pa}(V_{i}))$.
% 		Hence, $$V_{i} \notin \mathbf{M} \Rightarrow P_{\mathbf{m}}(V_{i} \mid \mathbf{pa}(V_{i})) = \mathcal{L}_{1}(V_{i} \mid \mathbf{pa}(V_{i}))$$
% 		Its contrapositive stands as well, $$P_{\mathbf{m}}(V_{i} \mid \mathbf{pa}(V_{i})) \neq \mathcal{L}_{1}(V_{i} \mid \mathbf{pa}(V_{i})) \Rightarrow V_{i} \in \mathbf{M}$$, which is the converse proposition of Case 1.
% 	\end{enumerate}

    In conclusion, $V_{i} \in \mathbf{M} \Leftrightarrow P_{\mathbf{m}}(V_{i} \mid \mathbf{pa}(V_{i})) \neq \mathcal{L}_{1}(V_{i} \mid \mathbf{pa}(V_{i}))$.
% 	Combine the two cases above, and we reach the conclusion
% 	$$V_{i} \in \mathbf{M} \Leftrightarrow P_{\mathbf{m}}(V_{i} \mid \mathbf{pa}(V_{i})) \neq \mathcal{L}_{1}(V_{i} \mid \mathbf{pa}(V_{i}))$$
\end{proof}

\section{Implementation Details}\label{sec:implementation-details}

\subsection{Baseline Methods}

Most of the code in this work is written in Python, while we adopt the R package pcalg~\cite{Kalisch:2012} for the PC algorithm.
We utilize process-based parallel programming to isolate errors only.

NSigma calculates $\max_{t} \frac{\lvert v_{i}^{(t)} - \mu_{i} \rvert}{\sigma_{i}}$.
We adopt the authors' implementation\footnote{https://github.com/Amossys-team/SPOT} for SPOT~\cite{Siffer:2017} while re-implementing ENMF~\cite{Cheng:2016} in Python based on the authors' MATLAB implementation\footnote{https://github.com/chengw07/CausalRanking}.
The other baseline methods are not publicly available.
We implement them by our understanding.

\subsection{Simulation Data Generation}

We generate the simulation datasets based on the Vector Auto-regression model, as shown in \eqnref{eqn:var}.
Following existing work~\cite{Zheng:2018,He:2021}, the value of non-zero elements in the weighted adjacent matrix, $A_{ij}$, is uniformly sampled from $(-2.0, -0.5) \cup (0.5, 2.0)$.
For the second item $\beta \mathbf{x}^{(t-1)}$, we set $\beta = 0.1$ in the experiment.
Finally, we sample the standard deviations from an exponential distribution for the zero-mean Gaussian noises $\mathbf{\epsilon}^{(t)}$.

The structure of $A$ is generated in two steps, as shown in \algref{alg:sedag-generation}.
We first generate a tree to ensure that the graph is a connected DAG.
Then, the other edges are inserted randomly.

\begin{algorithm}[htb]
	\caption{Graph Generation in the Simulation Study}\label{alg:sedag-generation}
	\begin{algorithmic}[1]
		\REQUIRE $N_{node}$, the number of nodes; $N_{edge}$, the number of edges
		\STATE $\mathcal{G} \gets (\mathbf{V}, \mathbf{E})$, where $\mathbf{V} = \{1, 2, \cdots, N_{node}\}$
		\FOR{$i = 2, \cdots, N_{node}$}
		    \STATE $j \gets$ choose one node from $\{1, 2, \cdots, i - 1\}$ randomly
		    \STATE Add the edge $i \rightarrow j$ into $\mathbf{E}$
		\ENDFOR
		\FOR{$k = N_{node}, N_{node} + 1, \cdots, N_{edge}$}
		    \STATE $i, j \gets$ sample $i, j \in \mathbf{V}$ randomly, \textit{s.t.}, $i > j \wedge (i \rightarrow j) \notin \mathbf{E}$
		    \STATE Add the edge $i \rightarrow j$ into $\mathbf{E}$
		\ENDFOR
		\RETURN $\mathcal{G}$
	\end{algorithmic}
\end{algorithm}

\subsection{Structural Graph Construction}

\begin{figure}[htb]
	\includegraphics{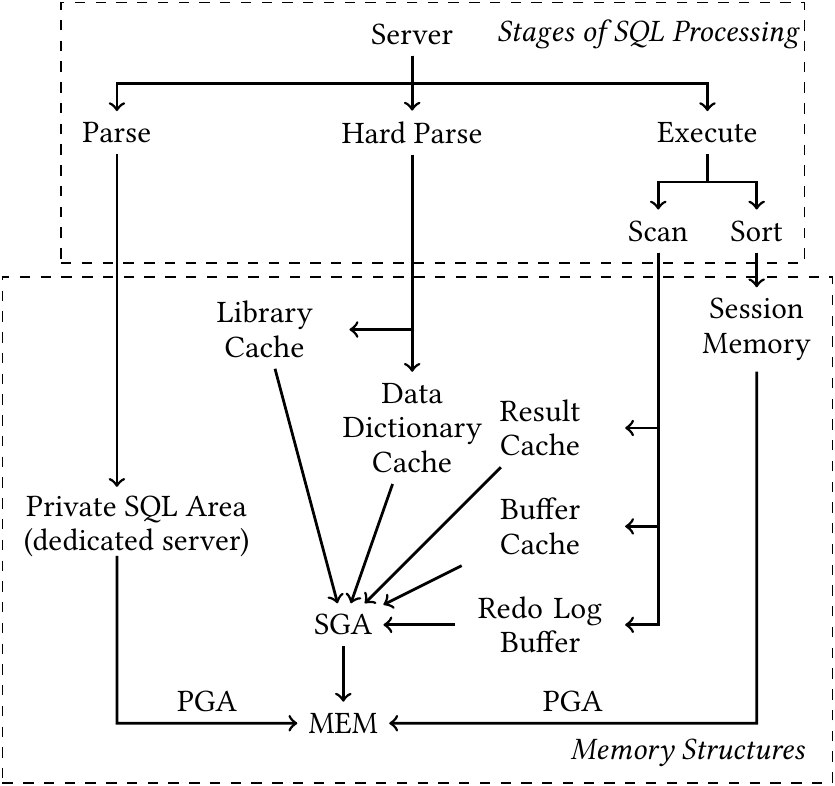}
	\caption{Part of the Oracle database call graph}\label{fig:oracle-architecture}
	\Description{
    	This figure shows the Stages of SQL Processing and Memory Structures.
    	We convert the related documentation into such a call graph.
	}
\end{figure}

In the empirical study, we construct the structural graph for the Oracle database instances.
\figref{fig:oracle-architecture} shows the SQL processing and memory structures in the call graph.
\figref{fig:performance:case} further shows part of the graph among metrics.
We drop metrics not included in the final structural graph, as our knowledge fails to cover them.
Baseline methods only use labeled metrics for a fair comparison.

\subsection{Regression Method Selection}\label{sec:regression-method-selection}

\tableref{tab:performance:oracle:regression} shows RHT's performance with several regression methods.
RHT with the linear regression (\textit{Linear}) has unsatisfying performance, as the relations among real-world variables are seldom linear.
Support Vector Regression (\textit{SVR}) with the sigmoid kernel, which is non-linear, improves the performance.
\citeauthor{Fu:2021} provide a way to predict distribution based on Random Forest (RF) and Mixture Density Networks (MDN), respectively, instead of a single value~\cite{Fu:2021}.
Hence, RHT combined with RF or MDN can measure the deviation for a new datum against the predicted distribution.
However, these two methods perform worse than the simple linear regression due to a limited understanding of the normal status, as shown in \figref{fig:ood}.
As a result, we choose SVR in the empirical study.

\begin{table}[tb]
	\caption{RHT with different regression methods on $\mathcal{D}_{O}$}\label{tab:performance:oracle:regression}
	\begin{tabular}{lrrrrr}
		\toprule
		\multirow{2}{0.18\columnwidth}{Regression Method} & \multirow{2}{*}{AC@1} & \multirow{2}{*}{AC@3} & \multirow{2}{*}{AC@5} & \multirow{2}{*}{Avg@5} & \multirow{2}{0.08\columnwidth}{T (s)} \\
		& & & & & \\
		\midrule
        Linear & 0.197 & 0.424 & 0.556 & 0.409 & 0.559 \\
        SVR & \textbf{0.328} & \textbf{0.601} & \textbf{0.677} & \textbf{0.546} & 0.834 \\
        RF & 0.202 & 0.394 & 0.525 & 0.382 & 22.065 \\
        MDN & 0.111 & 0.212 & 0.253 & 0.195 & 694.329 \\
		\bottomrule
	\end{tabular}
\end{table}

\end{document}